\documentclass[12pt]{article}

\usepackage{graphicx}
\usepackage{amssymb}
\usepackage{amsthm}
\usepackage{booktabs}
\usepackage{rotating}
\usepackage{multirow}
\usepackage{eurosym}
\usepackage{amsfonts}
\usepackage{amsmath}
\usepackage{umoline}
\usepackage{caption}
\usepackage{subcaption}
\usepackage{color}
\usepackage{url}
\usepackage{authblk}
\usepackage[english]{babel}

\usepackage{pbox}
\usepackage{footnote}
\usepackage{multirow}
\usepackage{changepage}
\usepackage{tablefootnote}
\usepackage{xcolor}

\usepackage[margin = 2cm]{geometry}
\usepackage{booktabs}
\usepackage{adjustbox} 
\usepackage{multirow} 
\usepackage{subcaption, epsfig} 
\usepackage{graphicx}
\usepackage{subcaption}
\usepackage{hyperref}

\usepackage{breakurl}
\usepackage{hyperref}
\hypersetup{breaklinks=true}

\usepackage[natbibapa]{apacite}

\usepackage{rotating}
\usepackage{lscape}
\usepackage{float}
\floatstyle{plaintop}
\restylefloat{table} 
\usepackage{booktabs}
\usepackage{textcomp}
\usepackage{siunitx}
\usepackage{multirow}
\usepackage{makecell}
\usepackage{longtable}
\usepackage{setspace}
\usepackage{amssymb}
\usepackage{amsmath}
\usepackage{booktabs}
\usepackage{adjustbox} 
\usepackage{multirow} 
\usepackage{subcaption, epsfig} 
\usepackage{graphicx}
\usepackage{subcaption}
\usepackage{hyperref}
\usepackage{utfsym}
\usepackage{multicol}
\usepackage{rotating}
\newcommand{\xmark}{\usym{2613}}
\usepackage{array}
\usepackage{natbib, ragged2e}
\usepackage{appendix}
\usepackage{pdflscape}
\usepackage{adjustbox}

\begin{document}
\newcolumntype{B}[1]{>{\small\hspace{0pt}\RaggedRight\bfseries}p{#1cm}}
\newcolumntype{C}[1]{>{\RaggedRight}p{#1cm}}

\title{Optimizing Credit Limit Adjustments Under Adversarial Goals Using Reinforcement Learning\footnote{\scriptsize NOTICE: This is a preprint of a published work. Changes resulting from the publishing process, such as editing, corrections, structural formatting, and other quality control mechanisms may not be reflected in this document. Changes may have been made to this work since it was submitted for publication. Please cite this work as follows: Alfonso-Sánchez, S., Solano, J., Correa-Bahnsen, A., Sendova, K. P., and Bravo, C. (2024). Optimizing credit limit adjustments under adversarial goals using reinforcement learning. \textit{European Journal of Operational Research} 315(2): 802-817, DOI: https://doi.org/10.1016/j.ejor.2023.12.025. This work is made available
under a Creative Commons BY license. \textcopyright CC-BY-NC-ND}}

\author[1, 2]{Sherly Alfonso-Sánchez}
\author[3]{Jesús Solano}
\author[3]{Alejandro Correa-Bahnsen,\footnote{Current affiliation: Anheuser-Busch InBev SA/NV. Cerrada De Las Palomas No.22, Piso 5. Reforma Social. Mexico City, 11650, Mexico.}}
\author[1]{Kristina P. Sendova}
\author[1]{Cristi\'{a}n Bravo}

\affil[1]{Department of Statistical and Actuarial Sciences, Western University, 1151 Richmond Street, London, Ontario, N6A 5B7, Canada.}
\affil[2]{Departamento de Matemáticas, Universidad Nacional de Colombia. Ave Cra 30 \#45-03, Edificio 404. Bogotá, Colombia.}
\affil[3]{Rappi, Cl 93 \#19-58, Bogotá, Colombia.}

\date{}
\maketitle
\begin{abstract}
Reinforcement learning has been explored for many problems, from video games with deterministic environments to portfolio and operations management in which scenarios are stochastic; however, there have been few attempts to test these methods in banking problems. In this study, we sought to find and automatize an optimal credit card limit adjustment policy by employing reinforcement learning techniques. In particular, because of the historical data available, we considered two possible actions per customer, namely increasing or maintaining an individual's current credit limit. To find this policy, we first formulated this decision-making question as an optimization problem in which the expected profit was maximized; therefore, we balanced two adversarial goals: maximizing the portfolio's revenue and minimizing the portfolio's provisions. Second, given the particularities of our problem, we used an offline learning strategy to simulate the impact of the action based on historical data from a super-app (i.e., a mobile application that offers various services from goods deliveries to financial products) in Latin America to train our reinforcement learning agent. Our results, based on the proposed methodology involving synthetic experimentation, show that a Double Q-learning agent with optimized hyperparameters can outperform other strategies and generate a non-trivial optimal policy not only reflecting the complex nature of this decision but offering an incentive to explore reinforcement learning in real-world banking scenarios. Our research establishes a conceptual structure for applying reinforcement learning framework to credit limit adjustment, presenting an objective technique to make these decisions primarily based on data-driven methods rather than relying only on expert-driven systems. We also study the use of alternative data for the problem of balance prediction, as the latter is a requirement of our proposed model. We find the use of such data does not always bring prediction gains.
\end{abstract}


\section{Introduction}
\label{sec:sample1}

All lending decisions imply evaluation of the capacity and motivation of a borrower to repay a loan and establishment of the necessary protection against losses due to default \citep{bazarbash2019fintech}. Indeed, forecasting the future behavior of borrowers has become a keystone for the credit management process. Some banking examples, according to \cite{hatzakis2010operations}, include forecasting involved in the decision to give credit to a new borrower (\textit{credit scoring}), forecasting the potential customer's likelihood to respond to an offer for credit (\textit{respond scoring}), or projecting the possibility of a current customer carrying a positive balance (\textit{balance scoring}). At present, the enhancement of methods used to generate the mentioned scores is a pivotal objective not only for traditional banks but also for financial technology companies, commonly referred to as \textit{fintechs}, and even some emerging \textit{``super-apps"}. These \textit{super-apps}, offer a range of services beyond just financial ones within a single environment (e.g., goods deliveries and social networks), and that them apart from conventional banks by their possession of \textit{alternative data} \citep{siddiqi2017intelligent}. This alternative data is generated by their corresponding applications and serves as a supplementary source to traditional financial data. 
The impact of incorporating this alternative data into decision-making models has been a growing area of research since the late 2000s \citep{djeundje2021enhancing}. Indeed, it has been demonstrated its effectiveness in improving predictive accuracy in some cases when compared to using only traditional data, in works such as,\cite{oskarsdottir2019value}, \cite{roa2021super} and, \cite{djeundje2021enhancing}.

In addition to the previous problems in general lending, financial institutions face more intricate decisions concerning revolving credit products (e.g., credit cards, home equity lines of credit, and personal and business lines of credit) compared to installment loans (e.g., personal loans and mortgages). In the case of credit card portfolios, additional to the interest-rate setting, retail banking companies must determine the initial credit limit for each customer and define the policy to adjust those limits over time. Because algorithms based on machine learning (ML) can detect complex patterns in data, presenting the opportunity for cost reduction and productivity improvement, they have been widely adopted in risk management \citep{leo2019machine}. According to the Basel III Accord, credit risk modeling mainly requires the estimation of the probability of default of every operation, which traditionally has been calculated using logistic regression \citep{thomas2017credit}. However, with the ML disruption, academic investigators and risk managers have tested different models looking for performance enhancements. In particular, \cite{lessmann2015benchmarking}
explored the influence of 41 classifiers in probability of default modeling across eight real-world data sets. Their findings showed that several classifiers could perform better than the industry-standard logistic regression in predicting credit risk. More recently, \cite{ala2021modelling} demonstrated that a bidirectional long short-term memory neural network could perform better than traditional methods such as gradient boosting \citep{tian2020credit}, bagging neural networks \citep{teles2020artificial}, support vector machine \citep{roy2019credit}, random forest \citep{rao20202}, and logistic regression \citep{bussmann2021explainable} in the context of calculating the probability of non-payment.

Academic researchers have looked to the deployment of ML models to address various tasks in the banking industry. For instance, to detect fraudulent events in credit card transactions in online purchases, \cite{van2015apate} have proposed APATE, a novel approach that combines features derived from incoming transactions and spending history using recency-frequency-monetary (RFM) variables (used also by \cite{khajvand2011estimating} and \cite{abbasimehr2022analytical}) and network-based features created by exploiting the dynamic social network of credit card customers and merchants. They found that employing all these features produced the best AUC (i.e., the area under the receiver operating characteristic curve) performance. For the same classification task, \cite{bin2022review} used a hybrid model based on neural-network-based models and federated learning concepts. In addition, \cite{katyayan2022analysis} used unsupervised ML models such as $K$-means and hierarchical clustering for customer segmentation. \cite{patil2019artificial} leveraged natural language processing (NLP) techniques for the creation of chat box interaction for client services purposes, and \cite{stevenson2021value} found that the use of text-based loan assessments without traditional data was effective in predicting default by employing recent advances of deep learning and NLP, specifically bidirectional encoder representations from transformers \citep{devlin2018bert}.

It is worth highlighting that most of the studies mentioned earlier fall within the domain of supervised learning (SL). In SL, we need a labeled dataset with pairs of inputs and outputs. The goal is to use statistical algorithms to learn a fixed mapping from input data to outputs. This is mainly driven by the desire to predict new outcomes based on new input observations \citep{jordan2015machine}. Based on this characterization, we can conclude that SL has a static nature. However, certain tasks necessitate sequential decision-making processes that find more appropriate frameworks in dynamic approaches, as opposed to traditional SL methods. The application of reinforcement learning (RL) is particularly apt for accommodating such dynamics. Unlike supervised learning, RL doesn't rely on labeled data; instead, it entails an agent interacting with an environment. The agent makes sequential decisions or takes actions, while the environment provides feedback in the form of rewards. The agent is driven by its aspiration to learn a policy that maximizes cumulative rewards over time. Consequently, the nature of RL is inherently dynamic.

Concerning RL, there have been few attempts in the state-of-the-art models \citep{singh2022reinforcement} to evaluate these techniques in credit risk management, in particular for the modification of credit lines. Although credit limit setting is an essential problem for traditional banking industries and Fintech companies, since identifying the adequate credit limit will define the profit and therefore the sustainability of the credit card portfolio, this question has not been widely studied. This contrasts with, for example, the default prediction problem, in which a large number of banking analytic research papers are published. Therefore, our study contributes to provide a data-driven methodology for the pressing challenge of credit line changes using the RL framework. In line with prior research, we embrace the concept of dynamic, automatic, approaches for formalizing credit line modifications. The reason behind this alignment is that any alteration to the credit limit for individual customers has an impact on the company's current financial standing. It directly influences the capital required to cover potential losses, often referred to as provisions, stemming from non-payments. These changes in provisions should be accounted for making subsequent decisions related to credit limit adjustments (to grant the company remains profitable and satisfies capital regulations), creating a sequential decision-making scenario. Hence, RL emerges as a suitable and intriguing framework for exploring credit limit setting. Additionally, in low-amount/high-growth strategies, such as the one the context this work operates in, credit limit increases are common and must be evaluated for all customers often. This means automatizing this process using RL is an attractive choice.

 Namely, the work described here experimented with an RL approach to credit limit adjustment on credit card portfolios which, unlike other stochastic scenarios where RL has been applied, such as investment portfolios that typically rely on online training, our approach requires off-line learning. For this purpose, we used information from a Latin American super-app, which has recently launched a credit card product. To offer more inclusion and enter the credit card market strongly, they have followed a low-amount, high-volume strategy by allowing a small credit limit for a significant proportion of customers and pursuing an aggressive strategy for limit increases for high-value customers. This fintech had thus far used an ad-hoc approach to change some customers' credit limits depending on their behavior to maximize its profits. Taking a comprehensive view of our work, to apply the RL technique, we require the construction of a simulator of the customer's responses after the decision about credit line changes is made. We aim to evaluate the impact of the inclusion of the given alternative data on the simulator model; subsequently, after developing the correct simulator we strive to train and compare the RL learning performance with other strategies, through synthetic experiments. 

With reference to the context mentioned above, our research was guided by three main objectives. Firstly, we aimed to define the essential components of an RL model capable of generating and automating optimal periodic recommendations for determining whether a customer should receive a credit limit increase. This involved framing the credit limit adjustment problem as an RL challenge. Additionally, given the necessity of constructing a simulator for offline training of RL algorithms for credit line modification, our simulator incorporates predictive models concerning the future types and amounts of balances. Therefore, our second research objective is to provide insights into the potential benefits of incorporating specific app usage variables into this simulator construction. This approach draws inspiration from previous studies, such as \cite{de2019does}, which examined the use of Facebook profile information to identify loan applicants who may struggle with loan repayment, and \cite{berg2020rise}, which explored digital footprint variables (e.g., device type, operating system, and email provider) to enhance traditional data for default prediction in the context of a furniture e-commerce company in Germany. In our case, we apply this concept to the relatively understudied task of revolving credit line balance prediction. Lastly, our final research objective involves comparing the performance of the policy generated by this RL technique against other strategies given our synthetic procedure. This comparative analysis could serve as a pivotal starting point, motivating further investigations into the application of this recent methodology in credit limit adjustment, mainly encouraging its evaluation in other real-world scenarios.

The remainder of this paper is structured as follows. In Section 2, we review previous work, and in Section 3 we introduce background context regarding RL and provisioning in credit card products. The formulation and motivation of the credit limit adjustment as an RL problem are presented in Section 4. Subsequently, Section 5 describes the experimental design, including details about the dataset, simulator construction, choice of RL algorithm, and specific aspects for comparing RL performance with other strategies. We present the results of the candidate models for the simulator and its interpretation and the results after training the RL algorithms in Section 6. In Section 7, we present the workflow for deploying this strategy. Finally, we present the key insights of the work in Section 8.

\section{Literature Review}
\label{sec:sample2}

\subsection{Modelling Historical Credit Card Limit Changes and Establishing Initial Credit Card Lines.}

There is a significant existing body of literature regarding modeling of credit limit adjustments based on historical information, without explicitly proposing a decision to change these limits depending on the new information of credit card holders' management acquired through time; instead, the purpose of these models is to deduce some structural facts about determinants of credit lines and response to credit line variations. For example, \cite{dey2009determinants} using data from the 1998 U.S. Survey of Consumer Finances, found statistical evidence that variables such as logarithm of income, credit rate, self-employment status, and age have a positive correlation with the given credit supply. And \cite{fulford2015important} characterized the credit card limit process as a stochastic one, with a Markov component that determines whether the customer can borrow and a separate component that determines the amount of the limit that the customer can borrow. This formulation is similar to what \cite{carroll1997nature} described for the modeling of income processes by using variables such as age, length of credit card tenure, geographical location, and the number of reported cards. Subsequently, they calibrated this model's parameters and explained that credit limit volatility affects how customers save, borrow, and spend, and provides the reason why one-third of American households present what is called the ``credit card puzzle", which is the phenomenon of carrying large revolving credit card debts at high-interest rates while holding liquid assets that pay low interest or none at all. Additionally, \cite{gross2000consumer} explored customers' responses to credit line modifications using data from a U.S. credit card portfolio, and their results showed that credit limit increases suggest a significant rise in debt, regardless of credit limit utilization; and more recently, \cite{aydin2022consumption}  conducted a controlled trial at a large Turkey retail bank, in which some customers (who had been preapproved for a credit line rise) were randomly selected to receive an increase of approximately 145\% of the post-tax income, and this empirical analysis showed that the shock in credit limits had a measured and significant effect on the use of credit, indeed it was found that borrowing rises by 11 cents per 1 Turkish lira of limit increase in the first quarter.

There are some standard ways to set the initial credit limit and the interest rate for credit card users, according to \cite{agarwal2018banks}, most credit card issuers use credit score models to decide the interest rate to be charged and the credit limit to be set for each credit card holder. For this purpose, banks and fintech companies generally develop their own internal models (involving analysis of the correlations between client characteristics, contract terms, and observed outputs in terms of profitability and default) and use them together with externally purchased credit scoring models, such as the FICO bureau score. Financial companies split their customers into groups based on their external credit scores and findings from internal procedures. Subsequently, a particular limit amount is set for each group, and lower credit limits are granted to customers with lower credit scores. 

\subsection{Strategies for Credit Limit Adjustment.}
Although setting the initial credit limit/interest rate is an important task for the credit card portfolio's constitution, there is no standard strategy that guides credit limit adjustment management in financial institutions. Indeed the current proposal methodologies are based on manual and expert interventions, or they tend to ignore the impact of expected losses (also known as provisions) due to default (e.g., \cite{fahner2012estimating}, \cite{ sohn2014optimization}), or they are supported under strong distribution assumptions (e.g. \cite{bierman1970credit}). For instance, \cite{so2011modelling} proposed a dynamic credit limit-setting policy generated from the use of behavioral scores and dynamic programming; they chose to capture customers' behavioral changes by incorporating the probability of default directly instead of using a static risk-return approach. However, this research did not explicitly show how the profit is calculated and if it is considering the provisions and the revolving nature of the credit card products in their calculation. \cite{charlier2019mqlv} introduced an RL algorithm called Modified Q-Learning for Vasicek (MQLV) to determine an optimal money-management policy for retail products based on clients' aggregate financial transactions, but they required the assumption that the transactional data sets satisfied the Vasicek model's assumptions. In these scenarios, they showed that the estimation of the probability of default generated by the MQLV algorithm provided more accurate results than using the Black-Scholes-Merton formula, which is well-known for portfolio management but was applied in this case for the administration of retail products. To provide additional context, Table~\ref{tab:Previous-studies} presents a summary of prior studies related to credit limit adjustments. The table outlines the framework employed in these studies, whether they considered relevant the provisions calculations (including the revolving nature of credit card products), if they addressed the dynamic nature of the problem, and whether the proposed methodologies underwent real-life testing.  Given this scenario we aim to contribute to the credit limit management literature, proposing a data-driven procedure to respond to the credit limit adjustment question with a profit focus by using the RL framework, which has shown to be a promising ML approach to solving decision-making problems.

\begin{table}[htbp]
\small
\begin{adjustbox}{max width=1\textwidth,center}
\begin{tabular}{@{}B{1.7}C{3.5}C{2.3}cccC{2.3}}

\hline
\textbf{}                                                          & \textbf{}                                                                                                                                                                                                                                                           & \textbf{}                                                                                                                                                                                                                           & \multicolumn{3}{c}{\textbf{Accounts for}}                                                                                                                                                                                                                                                                                                                                                           & \textbf{}                                                                                                                                              \\ \hline
\textbf{\begin{tabular}[c]{@{}l@{}}Authors \\ (Year)\end{tabular}} & \textbf{Framework}                                                                                                                                                                                                                                                  & \textbf{Data Set}                                                                                                                                                                                                                   & \multicolumn{1}{l}{\textbf{\begin{tabular}[c]{@{}l@{}}Provisions \\ (without \\ considering \\ revolving \\ characteristics \\ of credit cards)\end{tabular}}} & \multicolumn{1}{l}{\textbf{\begin{tabular}[c]{@{}l@{}}Provisions\\ (considering \\ revolving\\ aspect)\end{tabular}}} & \multicolumn{1}{l}{\textbf{\begin{tabular}[c]{@{}l@{}}Dynamic\\ nature \\ of the\\  problem\end{tabular}}} & \textbf{\begin{tabular}[c]{@{}l@{}}Results in \\ real-word \\ scenarios\end{tabular}}                                                                  \\ \hline
                                                   \citet{bierman1970credit}                & \begin{tabular}[c]{@{}l@{}}Dynamic  \\programming.\\ Strong Assumptions: \\ - Partial payments \\ \ \ no considered.\\ - The distribution of\\ \ \  collection probability\\ \ \ assumed to be beta.\end{tabular}                                                                         & \begin{tabular}[c]{@{}l@{}}None\\ The findings are \\exclusively\\ theoretical, \\lacking any\\ accompanying \\experimental\\ data.\end{tabular}                                                                                          & $\checkmark$                                                                                                                                                   & $\xmark$                                                                                                              & $\checkmark$                                                                                               & $\xmark$                                                                                                                                               \\ \hline
                                             \citet{so2011modelling}             & \begin{tabular}[c]{@{}l@{}}Markov decision \\ process with value \\ iteration algorithm.\\  Estimation of \\ transition \\ probabilities \\among \\ behavioral states.\end{tabular}                                                                                          & \begin{tabular}[c]{@{}l@{}}Credit card \\data from \\ major \\ Hong Kong bank.\end{tabular}                                                                                                                                              & \begin{tabular}[c]{@{}l@{}} No \\indicated \end{tabular}                                                                                                                                                  & \begin{tabular}[c]{@{}l@{}} No \\indicated \end{tabular}                                                                                                                 & $\checkmark$                                                                                               & \begin{tabular}[c]{@{}l@{}}$\xmark$\\ Comparison of \\the proposed \\methodology  \\with the \\ current one,\\ based on the \\model \\constructed.\end{tabular} \\ \hline
                                                                   \citet{fahner2012estimating} & \begin{tabular}[c]{@{}l@{}}\\Causal Learning \\using propensity \\ score matching.\\ - Counterfactual\\ \ \ prediction over\\ \ \  the balance response.\end{tabular}                                                                                                             & \begin{tabular}[c]{@{}l@{}}Credit card\\ accounts, \\ the specific \\data source\\  is not \\provided.\end{tabular}                                                                                                                       & $\xmark$                                                                                                                                                       & $\xmark$                                                                                                              & $\xmark$                                                                                                   & \begin{tabular}[c]{@{}l@{}}$\xmark$\\ The policy \\generated is \\ characterized \\in term of \\ income and risk \\levels.\end{tabular}                      \\ \hline
                                                                   \citet{sohn2014optimization} & \begin{tabular}[c]{@{}l@{}}-Regression models\\ \ \ to predict default \\ \ \ probability and \\ \ \ current balance with \\ \ \ the credit line as \\ \ \ input.\\ - Customer clustering \\  \ \ using regression trees\\ - Genetic algorithm to\\ \ \ find    the optimal \\ \ \ decision\end{tabular} & \begin{tabular}[c]{@{}l@{}}Data sourced \\from the \\ Strategy Designer \\Sample Files \\by Fair Isaac \\ Corporation (FICO)\end{tabular}                                                                                              & $\checkmark$                                                                                                                                                   & $\xmark$                                                                                                              & $\xmark$                                                                                                   & \begin{tabular}[c]{@{}l@{}}$\xmark$\\ \\ Results based on\\ the proposed \\methodology. \\ (Backtesting)\end{tabular}                                   \\ \hline
                                                                   \citet{charlier2019mqlv} & \begin{tabular}[c]{@{}l@{}}Modified Q-Learning \\ for Vasicek (MQLV)\\ Assumptions:\\ - Transactional \\ \ \ data satisfies Vasicek \\ \ \ model's  hypothesis.\end{tabular}                                                                                                   & \begin{tabular}[c]{@{}l@{}}Not publicly \\ release an anonymized \\ transactions dataset\\ due  to privacy, \\confidentiality, \\and regulatory \\ constraints. \\ (satisfies\\ Vasicek model \\assumptions).\end{tabular} & $\xmark$                                                                                                                                                       & $\xmark$                                                                                                              & $\checkmark$                                                                                               & \begin{tabular}[c]{@{}l@{}}$\xmark$\\ \\ Experimental \\ results  \\on the \\ risk estimation  \\rather than on \\ credit limit  \\modifications.\end{tabular}     \\ \hline
\end{tabular}
\end{adjustbox}
\caption{Summary of previous research on credit limit adjustment.}
\label{tab:Previous-studies}
\end{table}

\subsection{RL beyond deterministic scenarios: Applications in Business, Banking and Insurance.}
Overall, the main objective of an RL algorithm is to maximize a reward signal, achieved by discovering optimal actions through trial and error \citep{sutton2018reinforcement}. This stands in contrast to SL where the focus lies in learning a static mapping from input data to corresponding outputs (targets). These techniques have been applied in several contexts; for example, Deep Mind's AlphaGo RL agent \citep{silver2017mastering} was able to defeat the world champion of the game of Go in 2015. Several algorithms have been proposed in many articles to solve decision problems using these techniques, mostly in deterministic environments. For instance, \cite{mnih2015human} mixed deep learning and RL techniques to create a deep Q-Network agent that outperformed previous algorithms in the challenge of playing classic Atari games; this agent could achieve human-level control over 49 of these games.

As the complexity and success of RL algorithms have increased \citep{gronauer2022multi}, many academics and practitioners have begun to explore their application in stochastic environments. In the field of operations management, \cite{yang2022dynamic} presented an application of deep reinforcement learning (DRL) to generate an automated system to sell fresh products; they considered quality-based pricing and information disclosure. For this purpose, they used an automated cloud-based system to obtain quality measures and showed that this approach could reduce food waste and maximize the company's profit. \cite{ma2021monopoly} studied a pricing mechanism for online services using a traditional RL method (Q-learning) in which the objective was to learn about the quality of service provided to customers.\cite{yan2023online} developed a novel online model-based RL algorithm to optimize the electric vehicles (EV) scheme, aiming to determine which EVs to serve which orders and which EVs should be recharged, an important problem for the growing network companies that offer e-hailing services, such as Uber Green or DiDi Chuxing with more than 1 million EVs registered on its platform by 2021. \cite{de2022reward} used reward shaping (a technique that manipulates rewards) for policy transfer to encourage a DRL agent to learn from a teacher heuristic, instead of learning from scratch; this showed how analytical results could improve numerical methods in terms of performance, and stabilizing and accelerate the training process, they considered this strategy for solving the conventional perishable inventory problem, which is managing the inventory levels of perishable goods or products to optimize profits while minimizing waste or degradation. Also, several RL agents have been proposed for the management of investment portfolios. For example, \cite{wang2019alphastock} trained a DRL agent (Alpha Stock) capable of maximizing the portfolio's Sharpe value during a given period by considering the portfolio's risk-adjusted return. \cite{weng2020portfolio} constructed a DRL agent that observed the states given by the assets' daily low price, high price, and close price, which were the features selected as the most important variables after the application of an XGBoost model. In addition, they proposed the adoption of separable convolution and three-dimensional attention-gating networks to extract historical data features to allow their agent to outperform traditional algorithms.

With regard to the banking and insurance sectors, academic investigators and risk managers have attempted to support some decision-making processes through the application of RL. Such has been the case for the determination of the loan-acceptance threshold, adjustment of individuals' premiums in a car insurance portfolio, and the detection of fraud probability, which were formulated as RL problems by \cite{herasymovych2019using}, \cite{krasheninnikova2019reinforcement}, and \cite{el2017fraud}, respectively. These previous studies employed traditional RL algorithms, which did not include deep learning approaches to find the respective optimal policies.

\section{Background}
\label{sec:sample3}
\subsection{Reinforcement Learning}
\label{subec:subs31}
The mathematical foundation of an RL problem arises from a Markov decision process (MDP), which consists of a set of states $S$, an action space $A$, a reward function $r: S \times A \rightarrow \mathbb{R}$,  with $r(s, a)$ representing the numerical output given by the selection of action $a \in A$ when the observed state is $s \in S$, and a transition probability function $\mathbb{P}: S \times A \times S \rightarrow [0, 1]$, where $\mathbb{P}(s'\vert s, a)$ represents the conditional probability of transition to state $s'$ after the action $a$ has been taken in state $s$. Overall, if the MDP is episodic and involves discrete time steps, in each step $n$ the agent observes a state $s_n \in S$ and, based on $s_n$, selects an action $a_n \in A$. Subsequently, the agent receives a reward $r_n$ for the action $a_n$; after that, it observes the following state $s_{n+1}$. At step $n$, the agent's last objective is to find a policy $\pi:S \rightarrow A$ that maximizes the total expected discounted return $G_n$, according with the following equation:
\begin{equation}
    G_n = \displaystyle \sum_{k=n}^{N}\gamma^{k-n}r_k, \label{eq:TotalReturn}
\end{equation}
where $r_k$ is the reward at step $k$ and $\gamma \in [0, 1]$ is the discount factor. Larger values of $\gamma$ indicate that the agent is more interested in long-term rewards; if $\gamma = 0$, the agent is myopic because it only considers immediate rewards. 
 
To solve this optimization problem, the temporal difference (TD) learning \citep{sutton1988learning} attempts to estimate the return when the state is $s$ and the persuaded policy is $\pi$. In particular, the Q-learning algorithm \citep{watkins1989learning}, which is in the category of TD-learning and has been widely used, estimates the action-value function $Q_{\pi}(s, a)$, defined as the following:
\begin{equation}
    Q_{\pi}(s, a) =\mathbb{E}_{\pi}\left[G_n \middle| s_n =s, a_n=a\right]. \label{eqn:Qfunction}
\end{equation}
Intuitively, this function indicates the \textit{quality} of selecting the action $a$ in the state $s$. In addition, if $Q^*$ represents the optimal action-value function, the Bellman optimality equation states the following:
\begin{equation}
    Q^*(s, a) =\mathbb{E}_{\pi}\left[r_n + \gamma \max_{a_{n+1}}Q^*(s_{n+1}, a_{n+1})\middle| s_n =s, a_n=a\right], \label{eqn:Bellmaneq}
\end{equation}
and based on~\eqref{eqn:Bellmaneq}, the update equation in the Q-learning algorithm is given by:
\begin{equation*}
    Q_{\pi}(s_n, a_n) \leftarrow Q_{\pi}(s_n, a_n)+\alpha\left(r_n+\gamma\max_{a_{n+1}}Q_{\pi}(s_{n+1}, a_{n+1})-Q_{\pi}(s_n, a_n)\right), 
\end{equation*}
where $\alpha>0$ is the learning rate. To apply Q-learning, it is necessary to balance the exploration and the exploitation of the actions. Therefore, an $\epsilon$-greedy policy is introduced; that is, with $\epsilon$ probability, random action is selected, and with chance $1-\epsilon$, one action with the highest value of $Q$ is favored. This algorithm is categorized in the off-policy class because for the estimation of the $Q$ action value in the following state, one of the actions with the maximum value is considered rather than the output from the $\epsilon$-greedy policy.

It is important to note that the Q-learning algorithm, equation~\eqref{eqn:Bellmaneq}, includes the estimation of $\max_{a_{n+1}}Q_{\pi}(s_{n+1}, a_{n+1})$. This condition produces an overestimation problem because to estimate a maximum value, the greatest number of estimated values is employed, which can lead to a positive bias, also known as \textit{maximization bias} \citep{sutton2018reinforcement}. Therefore, the idea of Double Q-learning has emerged; instead of using the same Q-table to select the action with the maximum value and its corresponding estimate, the experiences are divided between learning two independent policies, $Q_1$ and $Q_2$. Without loss of generalization, $a^*$ is selected as the action that results in the maximum value within table $Q_1$; that is, $a^* = \arg \max_{a}Q_1(a)$. Next, $Q_2(a^*)$ is calculated by choosing the value within table $Q_2$ corresponding to the action $a^*$. To choose the policy used to the action selection, we sample from a Bernoulli distribution with $p=0.5$. If the Bernoulli sample number is $0$, the update rule is as follows:

\begin{equation}
    Q_1(s_n, a_n)\leftarrow Q_1(s_n, a_n)+\alpha\left[r_n+\gamma \  
 Q_2\left(s_{n+1},\arg \max_{a}Q_1(s_{n+1},a)\right)-Q_1(s_n, a_n)\right], \label{eqn:DoubleQ}
\end{equation}

In the other case, if the Bernoulli sample number is $1$, $Q_2$ is exchanged with $Q_1$ in~\eqref{eqn:DoubleQ}. More details regarding these algorithms, have been presented by \cite{sutton2018reinforcement}.

\subsection{Provisioning in credit card products}
 
Despite the lack of common regulation, just like traditional banks, fintech companies must also be able to quantify credit risk in their portfolios to set provisions, calculated as the expected loss derived from unpaid card balances, and thus protect the company's profitability and financial stability. Moreover, this provision estimation should be done periodically to reflect changes in the customers' behaviors and new information available. Given that any limit change will bring about a change in credit risk, considering the change in provisions has to be a must. In general, credit card providers will have more profit if the customers have high outstanding balances, however customers with bad payment behavior and higher debts will represent potential costs to the company from provisions alone (even if no arrears or very limited arrears occur) and since the balances are dependent on the credit limit, adequate decision over the latter is crucial. Therefore to answer about credit limit adjustment the tradeoff between revenue and provisions has to be considered. For provision calculations, so far fintechs have adopted banking regulations, particularly the accepted methodology to calculate capital requirements according to the Basel III Accord. Under this regulation, the provision or expected loss, $\mathbb{E}(Loss)$ which represents an amount that must be kept in liquid assets to cover expected losses, is calculated using the concepts of the probability of default ($PD$), that is the likelihood that a borrower fails to make the required payments, the loss given default ($LGD$), or the percentage of the exposure at default that would be unrecoverable; and the exposure at default ($EAD$) which represents the total amount that the company expects to lose when default occurs. Finally, the expected value of the loss (the provision) is calculated as:

\begin{equation}
    \mathbb{E}(Loss_i) = PD_i\cdot LGD_i \cdot EAD_i,\label{eqn:Provision}
\end{equation}

\noindent where $\mathbb{E}(Loss_i)$, $PD_i, LGD_i$ and $EAD_i$ represent the expected loss, default probability, loss given the default and the exposure at default, respectively, for the $i^{th}$ customer. In contrast to fixed-income exposure loans, such as consumer loans where the $EAD$ is calculated in a straightforward manner using the current outstanding balance, for revolving products such as credit cards, this approach does not reflect the fact that a borrower near default may spend more of their current remainder credit limit. For this reason, the utilization of the credit conversion factor (CCF), which includes the proportion over the remainder credit limit that is likely to be spent in the case of default, is mandated by the regulation. Specifically, if $L_i$ is the current credit limit for the $i^{th}$ customer:

\begin{equation}
    EAD_i = \textit{Outstanding balance}_i + CCF\cdot (L_i-\textit{Outstanding balance}_i). \label{eqn:EAD}
\end{equation}

As described by \citet{tong2016exposure}, one manner to estimate the CCF  of a portfolio is considering the information of the defaulters, if the total number of them is $D$, for each defaulter $d$, $d \in \{1, \ldots, D\}$, $CCF_d$ denotes the individual $CCF$. The $CCF_d$, then, is  defined as follows:

\begin{equation}
    CCF_d = \frac{OB(t_{def})_d-OB(t_r)_d}{L_d-OB(t_r)_d}, \label{eqn:CCFd}
\end{equation}

where $OB(t_{def})_d$ is the outstanding balance at the moment of default and $OB(t_r)_d$ and $L_d$ are the outstanding balance and the current credit limit at the beginning of the billing period in which the default occurred, respectively, for defaulter $d$. The numerator of~\eqref{eqn:CCFd} indicates the increase in the borrower's debt during the period in which the default event took place. This difference is then divided by the subtraction between the credit limit and the outstanding balance at the beginning of the period in which the default event happened. Therefore, the $CCF_d$ shows the proportion of the borrower's outstanding balance that is at risk of being lost in the event of default. Finally, the $CCF$ of the portfolio is set equal to the average of the individual $CCF_d$ factors.

\section{Limit Setting as a Reinforcement Learning Problem}
\label{sec:sube41}
We believe that previous approaches used to determine optimal credit limit adjustment policies can be enhanced in several ways. Firstly, when these approaches treat the problem conventionally, considering credit line decisions solely based on customer input features, they tend to overlook crucial real-life factors. These factors include the number of customers and the total amount provisioned for potential losses by the company (global state) before any credit line changes are made. Additionally, they disregard the dynamic nature of the company's financial situation after each decision, which impacts future credit line adjustments for other customers. For example, a significant increase in the company's provisions may necessitate more stringent actions. Therefore, we argue that framing credit line modifications as a sequential decision-making process, using RL in particular, is more apt. Secondly, compared to previous dynamic approaches like dynamic programming, RL learns an optimal policy from interaction with the environment (or its simulator) through trial and error. This means that we don't need to know in full the environment's dynamics, such as the transition probabilities among states, which can  be often unknown in real-life contexts. Thus, the RL framework effectively accommodates critical aspects of credit line adjustment, considering its sequential decision-making nature and offers flexibility due to its exploration and exploitation learning nature. In this section, we define credit limit setting as an RL problem, explicitly outlining its components, including states, actions, and reward functions. Particularly, this explicit exposition of the reward function includes terms accounting for the revolving characteristics of credit card products which has not been done in previous works.

Now for our experiments, according with the available data for this study, in which almost all the clients who have received an increase in their current credit limit obtained a constant factor $\beta$ of raise, we could not observe changes in behavior due to different factors of increase. In addition, as our goal is to propose a data-driven methodology based on the historical information of customers' behavior to credit limit adjustment, other  questions such as the determination of the interest rate have already been determined. With this context, we first defined our particular problem as one of RL; specifically, we had to decide to which customers to offer an increase $\beta>0$ factor over their current credit limit (i.e. the new limit will be $\beta$ percent greater than the current limit). Considering the expected profit of this offer, this is recognizing the expected revenue and provisions. We also analyzed the impact of the inclusion of alternative (app) data, such as RFM segment, number of registered addresses, and wealth index, in the construction of models that were used for the environment simulator, this is the model for emulating the response given the increase or maintenance of the credit limit.

Therefore, the set of available actions was $A = \{0, 1\}$, where $a=1$ represents an increase in the current credit limit by a factor $\beta$ (the company sets this factor based on their internal policies) and $a=0$ indicates that the current limit is not changed. Similarly to \cite{krasheninnikova2019reinforcement}, we formulated the problem as an episodic one, with the number of steps equal to the total number $N$ of credit card customers who possess 3 or more months of seniority and who at the time of decision making are up to date in payments. We chose this formulation as it would be ineffective to offer a limit increase to a deteriorating customer. At least 3 months of history were needed to estimate future behavior correctly. For each credit card holder's decision, we examined the customer's attributes retrospectively, specifically analyzing data from three months preceding the decision. This retrospective window was utilized for determining whether to increase or maintain their credit limit. Subsequently, we assessed the consequences of this decision by evaluating the customer's response at the end of the trimester immediately following the decision (prospective window).

\newpage
Below, we present and define the elements in the RL formulation:
\begin{itemize}
    \item \textbf{State ($s_n$)}: was defined in terms of not only credit card holders' characteristics but also in connection with the company provisions change given the credit line modifications. Although, in general, independence among customers is assumed, the credit limit adjustment will affect the total provisions necessary for the company. Thus, we first denote the historical customer's financial features $fc_n$ obtained from their  credit card usage as the following: 
      $$fc_n=(\overline{UR}_n, \overline{PR}_n, \overline{CR}_n, MP_n, L\_R_n, Int_n),
    $$
   
    where $\overline{UR}_n$, $\overline{PR}_n$, and $\overline{CR}_n$ are the monthly average, utilization, payment, and spending consumption rates, respectively, during the last 3 months for the $n^{th}$ customer. In addition, $MP_n$ represents the number of non-payments, $L\_R_n$ the current limit, and $Int_n$, the annual interest charged, in the last quarter. Furthermore, the state of the environment before making the $n^{th}$ decision is defined as the following:
    \begin{align}
        s_n = \left(fc_n, \ \Delta Provisions_{n-1}\right), \label{eqn:state}
    \end{align}
    \noindent where $\Delta Provisions_{n-1}$ is the total provisions' change given all the actions taken for the  $1^{st}, \ 2^{nd}, \ldots, \ (n-1)^{th}$ customer, compared with the provisions when the credit limits are maintained for all the customers in the portfolio, this last component is important due to, primary it represents the aggregate change in provision terms given the actions decided for the first $n-1$ customers of the portfolio and secondly it also allows that a next state is affected by the action taken in the previous state, a requisite to our problem to be framed as an RL one. For the initial state, as no action has been made, the following is set:
    \begin{align*}
        s_1 = \left(fc_1, \ 0\right).
    \end{align*}
    Subsequently, after making the first decision $a_1$, for the first customer, the environment state changes to the following:
    \begin{align*}
        s_2 = \left(fc_2, \ \Delta Provisions_1\right),
    \end{align*}
       where if $Provision_i(a)$ represents the provision required for the $i^{th}$ credit card customer when $a \in \{0, 1\}$ is the selected action, then $\Delta Provisions_1= Provision_1(a_1)+\displaystyle\sum_{i=2}^{N}Provision_i(0)-P_0$, with $P_0=\displaystyle\sum_{i=1}^{N}Provision_i(0)$. We note that $P_0$ is the provision given no changes in the current portfolio; therefore $\Delta Provisions_1= Provision_1(a_1)-Provision_1(0)$, which represents the change in provisions if the action for the first customer is $a_1$. In general, for the $n$ step, that is, before deciding whether the current limit for the  $n^{th}$ customer should be increased, the environment state is indicated by~\eqref{eqn:state}, in which $\Delta Provisions_{n-1}= \displaystyle \sum_{i=1}^{n-1}\left[Provision_i(a_i)-Provision_i(0)\right]$, where the actions taken for the customers $1, \ldots, \ n-1$ were $a_1, \ldots, \  a_{n-1}$, accordingly. $\Delta Provisions_{n-1}$ stands for the aggregate change in provisions given the actions until the $(n-1)^{th}$ customer.
    \item \textbf{Reward function ($r_n$)}: To define this function, we consider the expected profits during the following 3-month period after the company has selected the actions. In this case, the expected profit $\mathbb{E}(Profit_n\vert a_n)$ in step $n$ after performing the action $a_n$ is the difference between the expected revenue and the provisions that will be needed, as follows:
   \begin{align}
    \mathbb{E}(Profit_n\vert a_n) &= 3\cdot Int_n\cdot\bar{R}(a_n)\cdot(1-PD_n)-Provision_n(a_n), \label{eqn:EProfit}   
    \end{align}
   where using~\eqref{eqn:Provision} and~\eqref{eqn:EAD}, the following can be derived:
    \begin{align*}
    Provision_n(a_n) & = PD_n \cdot LGD \cdot \left[OB_{3,n}+CCF \cdot (L(a_n)-OB_{3,n})\right],
    \end{align*}
and with the following designations:
    \begin{itemize}
    \item[i)] $Int_n$ is the monthly interest rate charged to the $n^{th}$ customer.
    \item[ii)] $\bar{R}(a_n)$ is the predicted monthly average outstanding balance at the end of the prospective window for the $n^{th}$ customer when $a_n$ is decided.
    \item[iii)] $PD_n$ is the default probability of the $n^{th}$ customer calculated when making the decisions at the end of the retrospective window. This probability is calculated for each customer using a local regulation model, that depends on several elements, for instance, percentage of, payment, usage, and the number of no payments with the credit card.
    \item[iv)] $OB_{3,n}$ is the outstanding balance of the $n^{th}$ customer at the payment date starting the prospective window.
    \item[v)] $L(a_n)$ is the credit limit after performing the action $a_n$.
    \item[vi)] $CCF$ is the credit conversion factor of the credit card portfolio. 
    \end{itemize}
    
In equation~\eqref{eqn:EProfit}, the expected revenue is indicated by $3\cdot Int_n\cdot\bar{R}(a_n)\cdot(1-PD_n)$, which is estimated as three times the monthly interest amount perceived multiplied by the probability of not defaulting. This is due to the fact that our evaluation is done with a look-forward period of three months. Similar expressions are used in actuarial reasoning, namely for life contingencies, where for the calculation of a policy profit at the time of inception, the expected revenue and provisions are commonly called expected premiums and reserves, respectively \citep{dickson2019actuarial}.

To specify the reward function, we established the baseline strategy maintaining the current limits for all the customers. We selected this baseline strategy because it represented the company's \textit{status quo}, which we aimed to outperform through the policy generated by the RL algorithm. Specifically, $R_0$ denotes the portfolio's expected profit following this baseline strategy, as follows:
\begin{equation*}
R_0 = \displaystyle{\sum_{i=1}^N \mathbb{E}(Profit_i \vert a_i=0)},
\end{equation*}
where $N$ is the total number of customers in the credit card portfolio and $\mathbb{E}(Profit_i \vert a_i=0)$ is the expected profit of the $i^{th}$ customer for $i \in \{1, \ldots, N\}$, given no changes in credit limits. After the first action $a_1$ has been selected, that is, to increase or maintain the credit limit for the first customer; we define $R_1$ as:
\begin{equation*}
R_1 = \mathbb{E}(Profit_1 \vert a_1)+\displaystyle{\sum_{i=2}^N \mathbb{E}(Profit_i \vert a_i=0)}. 
\end{equation*}
If the reward in step 1 (i.e., after taking the action for the first customer) is determined as $r_1=R_1-R_0$, then, recursively, the reward in the $n^{th}$ step is specified by:
\begin{equation}
r_n = R_n-R_{n-1}, \label{eqn:Recursivereward}
\end{equation}
where $R_n = \displaystyle{\sum_{i=1}^n \mathbb{E}(Profit_i \vert a_i)}+\displaystyle{\sum_{i=n+1}^N \mathbb{E}(Profit_i \vert a_i=0)}$. From~\eqref{eqn:Recursivereward}, we can establish the reward function simply as the following:
\begin{equation}
    r_n = \left\{
	\begin{array}{ll}
		\mathbb{E}(Profit_n \vert  1)-\mathbb{E}(Profit_n \vert  0)  & \mbox{if }  a_n = 1, \\
		0 & \mbox{if } a_n = 0.
	\end{array}
\right. \label{eqn:Simplyreward}
\end{equation}
As shown in~\eqref{eqn:Simplyreward}, the reward function indicates the advantage in expected profit terms after having allowed a credit increase over having not increased the limit for each customer at the end of the next quarter.
\end{itemize}
In the previous setting, the order in which the customers are examined could influence the updating process, for this reason, to avoid any bias because of the order in which the data is inspected, we permuted the observations in every episode during the RL training stage.

\section{Experimental Design}
\label{sec:sample4}
Because of the specifications of our problem, acting online to store real-time experiences and train our RL algorithms was infeasible; first, increasing or maintaining credit limits to customers in a seemingly random manner would provide a very poor customer experience and would lead to poor customer management. Second, our reward is not immediate; instead, we can measure only the impact of our actions three months later. Consequently, we needed to train the RL algorithm in an off-line way unlike, for example, applications on portfolio investment using RL algorithms in which online training has been employed \citep{schnaubelt2022deep}.  Moreover, in contrast with other scenarios in which common resources such as the widely used OpenAI Gym Python library \citep{brockman2016openai}, now moved to the Gymnasium Python library \citep{towers_gymnasium_2023}, are used to investigate or study RL algorithms \citep{chen2021decision, ravichandiran2018hands}. Within these approaches the environments’ dynamics are already constructed as they are directly derived from the application itself. We require constructing a simulator of the real-world business environment with which our RL algorithms interact.  This construction represented the second challenge in our research after formulating the RL setup, it enabled us to generate appropriate models for emulating responses following the maintenance or increase of the current credit limit, using historical data. In the following subsections, the data used for this purpose and the ideas for the simulator construction are presented.
\subsection{Data set}
Rappi, a Latin American super-app, serves as an intermediary between businesses and customers through its mobile application, RappiApp. Due to its extensive reach and growth, Rappi has also ventured into providing financial services. The data set used in this research comes from the company Rappi and it is related to some of its credit card (RappiCard) holders who were active (not blocked and who had not canceled their product) from June to December 2021 in a Latin American country. This dataset comprises both financial and generic attributes of these customers.

The first set of attributes encompasses variables commonly recorded by credit card companies, which allow us to gain insights into the specific financial behavior of each client with respect to their credit product. Among these financial features are total credit card spending, payment history, outstanding balances, instances of missed payments, bureau credit scores, and whether the customer received a credit limit increase. The second set of attributes called generic features, is information associated with the usage of the RappiApp. These attributes provide insights into the customer's profile and preferences in terms of services and payment methods. Within this category, we find the number of orders made through RappiApp using the credit card, preferred product categories (e.g., restaurants, e-commerce, travel), and favored payment methods (e.g., credit card, cash). In Table~\ref{tab:Features-Description}, the definition of some selected features is provided. 

We employed observations of 12,258 customers who were up to date at the end of August 2021 to train the models for the simulator, while we used information regarding 11,101 customers who were up to date at the end of the prospective window to train the RL algorithms. Some of the differences in means between the data set used for RL training versus the data set used for the simulator constructions are: 15.79 USD, 35.15 USD, 29.75 USD, for the total consumer spendings for months 1, 2, and 3, respectively. Similarly, outstanding balances at the cut-off date of months 1, 2, and 3 display variations at 39.21 USD, 40.21 USD, and 46.06 USD, correspondingly. Additionally, payments at the payment date concerning the outstanding balance at the cut-off date for months 1, 2, and 3 show discrepancies of 19.33 USD, 14.55 USD, and 28.19 USD, respectively. With regards to the difference in means for the credit limits in the retrospective window and the bureau scores, the values are 50.66 USD and 0.41, in that order. We can observe that, on average, there is a slight increase in total spending consumption, outstanding balances, and payment amounts after credit line increases. However, these increases are relatively small, with the maximum increase being less than 47 USD. As expected, the average credit line for the dataset used in RL training is higher than that for the dataset used in constructing the simulator. Nevertheless, the difference in bureau scores between the two datasets is less than one unit.

\begin{table}[htbp]
\centering
\setstretch{1}
\resizebox{\textwidth}{!}{
\begin{tabular}{ll}
\toprule
\textbf{Feature}&\textbf{Description}\\\midrule
\begin{tabular}[c]{@{}l@{}}Financial \\Transactional\\ Data\end{tabular}                                                         & \begin{tabular}[c]{@{}l@{}}1.  \textbf{\texttt{TCi}}: Total consumer spending in the month \texttt{i}, for \texttt{i=1, 2, 3}, i.e., in the retrospective window.\\\\ 2. \textbf{\texttt{EO\_i}}: Operative state in the month \texttt{i},  which represents the consecutive number of non-payments \\ 
 \ \ \ \ \ \ \ \ \ \ \ \ \ \ \ until month \texttt{i}, for \texttt{i=1, 2, 3}.\\\\ 3. \textbf{\texttt{L\_R}}: Credit limit in the retrospective window.\\\\ 4. \textbf{\texttt{MP\_R}} : Total number of non-payments in the retrospective window.\\\\ 5. \textbf{\texttt{N\_Months\_R}}: Number of months with the credit card until the end of the retrospective period.\\\\ 6. \textbf{\texttt{Int}}: Annual interest charged to the credit card.\\\\ 7. \textbf{\texttt{EI}}: Estimated income.\\\\ 8. \textbf{\texttt{HA\_P}}: Binary variable to represent whether the customer had received an increase (\texttt{1}) or had not (\texttt{0}).\\\\ 9. \textbf{\texttt{L\_P}}: Credit limit set for the prospective window.\\\\ 10. \textbf{\texttt{OB\_cday\_i}}: Outstanding balance at the cut-off date of the month \texttt{i}, for \texttt{i=1, 2, 3}.\\\\ 11. \textbf{\texttt{P\_pday\_i}}: Payment at payment date with respect to \textbf{\texttt{OB\_cday\_i}}, \texttt{i}, for \texttt{i=1, 2, 3}.\\\\
12. \textbf{\texttt{BS}}: Bureau score at the moment of the application of the credit card.

\end{tabular} \\\\ \hline
\begin{tabular}[c]{@{}l@{}}Generic \\ Data \end{tabular} & \begin{tabular}[c]{@{}l@{}} \\ 1. \textbf{\texttt{AGE\_IN\_APP}}: Age in the app.\\\\ 2. \textbf{\texttt{N\_ORDERS\_CC}}: Number of orders until the date on which the data was extracted.\\\\ 3. \textbf{\texttt{SEGMENT\_RFM}}: Segment of the customer according to recency, frequency, and monetary value.\\\\ 4. \textbf{\texttt{N\_ADDRESS}}: Number of different addresses registered with app.
\\\\ 5. \textbf{\texttt{NATIONAL\_CC}}: Number of national credit cards.
\\\\ 5. \textbf{\texttt{INTERNATIONAL\_CC}}: Number of international credit cards. 
\\\\ 6. \textbf{\texttt{WEALTH\_INDEX}}: Index of   wealth.
\\\\ 7. \textbf{\texttt{FAV\_VERTICAL}}: Preferred product types.
\\\\ 8. \textbf{\texttt{FAV\_PAYMENT\_METHOD}}: Favorite payment method.
\end{tabular}                \\
\bottomrule
\end{tabular}}
\caption{Descriptions of some selected features. Generic data refers to information generated by the customer's use of the app.}
\label{tab:Features-Description}
\end{table}

\subsection{Simulator construction}
\label{subsec:Simulator_construction}
In this stage, our goal was to find an accurate model to predict the future monthly average outstanding balance at the payment date during the following 3 months. For this purpose, historical decisions that contained information regarding the 3,290 credit limit increments were considered. Because of the complex characteristics of the credit card customers' portfolio, using a single model did not provide an adequate solution for this problem. Indeed, many customers had remained inactive or had paid their outstanding balance in full (\textit{full payers}). Therefore, the number of zero values in the target variable was high. After various experiments, we found that using a two-stage model consisting of one classifier model followed by a regressor model was an adequate strategy. We first classified the balance type among three options and subsequently regressed the remaining outstanding balance, given the predicted balance type.

\noindent First, we classified customers across three classes depending on their balance type. In particular, we defined class 0 to be those customers who had a monthly average balance greater than 0 but less than or equal to 75.81 USD at the payment date; class 1 represented customers with a monthly average outstanding balance greater than 75.81 USD, and class 2 contained customers who had remained inactive or had paid their balances in full in the following quarter. We determined the previous cut-off of 75.81 USD for defining the balance types by exploiting prior analysis and selecting the one that results in the model with the best cross-validation error.

We evaluated different models for the classifier selection, including decision trees, random forest, and XGBoost; in addition, we examined the impact of inclusion of alternative features versus using only financial information. Because the proportions of the classes were 17\%, 26\%, and 57\% for classes 0, 1, and 2, respectively, we used SMOTENC, which is a variation of the Synthetic Minority Over-Sampling Technique \citep{chawla2002smote} that can be applied for nominal and continuous variables. SMOTENC oversamples the minority classes to tackle the problem of imbalanced classification; in particular, the technique generates continuous features of new samples using $k$-Nearest neighbors and interpolation, while for categorical features, the most frequent categories among the nearest neighbors are used. In addition, because of the imbalanced nature of the data set, during the hyperparameter search we used the weighted $f1$ score as the evaluation measure because it considered the weighted average of the $f1$ score for each class. The results of various model candidates for the simulator are given in the subsection~\ref{subsec:Simulator results}.

\subsection{RL algorithm and performance comparison}
After having the simulator constructed, we used it for interacting with the RL algorithm in an off-line way. With regards to the RL algorithm, given that the credit card portfolio was still emerging, we did not count with a large number of experiences about the customers' response given a credit line modification, thus similar to \cite{herasymovych2019using} and \cite{krasheninnikova2019reinforcement}, we opted for using traditional RL algorithm, namely the Double Q-learning (explained in subsection \ref{subec:subs31}). We performed hyperparameter search to find the best Double Q-learning agent and subsequently, we compared its learning process with other baseline policies, we emphasize that this comparison was done under synthetic experimentation given the proposed methodology. Finally, we provided an interpretation of the policy generated by the RL agent. The results of these procedures are found in the subsection~\ref{subsec:RLtraining_result}.

\section{Results}
\subsection{Impact of Incorporating Alternative Data on Simulator Construction and Its Interpretation}
\label{subsec:Simulator results}

According with the experimental design described in~\ref{subsec:Simulator_construction}, first we obtained the results regarding the classification task, this is the prediction of the balance type. Using both sets of predictors, namely alternative and financial features versus only financial features, the XGBoost model exhibited the best performance. We also observed that the inclusion of alternative features did not produce a significant improvement in the validation measure (only a 1\% rise in the weighted $f1$ score), as shown  in Table~\ref{tab:ModelSelection}.

\begin{table}[htbp]
\centering
\setstretch{1}
\begin{tabular}{lcccc}
\toprule
                    & \multicolumn{4}{c}{\textbf{Weighted f1-score}}
            
                    \\ \cmidrule{2-5} 
\multicolumn{1}{c}{\multirow{2}{*}{\textbf{Best}}} & \multicolumn{2}{c}{\textbf{
\begin{tabular}[c]{@{}c@{}}Alternative and Financial\\ predictors\end{tabular}}} & \multicolumn{2}{c}{\textbf{\begin{tabular}[c]{@{}c@{}}Only Financial \\ predictors\end{tabular}}} \\ \cmidrule{2-3} \cmidrule{4-5} 
\multicolumn{1}{c}{}                               & \multicolumn{1}{c}{\textbf{CV-score}}               & \multicolumn{1}{c}{\textbf{Test-score}}              & \multicolumn{1}{c}{\textbf{CV-score}}                    & \textbf{Test-score}                    \\ \midrule
\textbf{Decision Tree}                               & \multicolumn{1}{c}{0.61 $\pm$ 0.04}                            & \multicolumn{1}{c}{0.55 $\pm$ 0.02}                             & \multicolumn{1}{c}{0.60 $\pm$ 0.03}                                 & 0.55 $\pm$ 0.02                                   \\ 
\textbf{XGBoost}                                     & \multicolumn{1}{c}{\textbf{0.77 $\pm$ 0.14}}                            & \multicolumn{1}{c}{0.66 $\pm$ 0.01}                             & \multicolumn{1}{c}{\textbf{0.76  $\pm$ 0.09}}                                 & 0.65 $\pm$ 0.01                                 \\ 
\textbf{Random Forest}                               & \multicolumn{1}{c}{0.63 $\pm$ 0.05}                            & \multicolumn{1}{c}{0.58 $\pm$ 0.01}                             & \multicolumn{1}{c}{0.63 $\pm$ 0.02}                                 & 0.57 $\pm$ 0.01                                   \\ \bottomrule
\end{tabular}

\caption{Scores for model selection in the classification stage. 95\% confidence intervals are presented.}
\label{tab:ModelSelection}
\end{table}

\noindent Second, given the classification of balance type, we found that the best models to apply for class 0 and class 1 were XGBoost and random forest, respectively. We selected these regression models using root mean squared error (RMSE) measure, and for evaluation, we used weighted absolute percentage error (WAPE)\footnote{We employ WAPE since it accounts for the error based on the total amount of outstanding balances. $WAPE=\dfrac{\sum_i\vert y_i -\hat{y}_i \vert}{\sum_i \vert y_i\vert}$, where $y_i$ is the observable output and $\hat{y}_i$ its corresponding prediction.} measure. After training over both sets of predictions, we obtained the results shown in Table~\ref{tab:Regression_results}.

\begin{table}[htbp]
\centering
\setstretch{1}
\begin{tabular}{lcccc}
\toprule
                    & \multicolumn{2}{c}{\textbf{RMSE (USD)}}
                    & \multicolumn{2}{c}{\textbf{WAPE}}
                    \\ \cmidrule{2-5} 
 & \multicolumn{1}{c}{\textbf{\begin{tabular}[c]{@{}l@{}}Alternative \\ and Financial\\ predictors\end{tabular}}} & \multicolumn{1}{c}{\textbf{\begin{tabular}[c]{@{}c@{}}Only Financial \\ predictors\end{tabular}}}
& \multicolumn{1}{c}{\textbf{\begin{tabular}[c]{@{}l@{}}Alternative \\ and Financial\\ predictors\end{tabular}}} & \multicolumn{1}{c}{\textbf{\begin{tabular}[c]{@{}c@{}}Only Financial \\ predictors\end{tabular}}}
\\ \midrule
\textbf{\begin{tabular}[c]{@{}c@{}}Class 0 \\ Best XGBoost\end{tabular}}                              & \multicolumn{1}{c}{22.09 $\pm$ 1.02}                            & \multicolumn{1}{c}{\textbf{22.13 $\pm$ 1.08}}                             & \multicolumn{1}{c}{0.64 $\pm$ 0.04}                                 & 0.64 $\pm$ 0.04                                  \\ 
\textbf{\begin{tabular}[c]{@{}c@{}}Class 1\\ Best Random \\ Forest\end{tabular}}                                     & \multicolumn{1}{c}{209.32 $\pm$ 23.35}                            & \multicolumn{1}{c}{\textbf{192.30$\pm$ 22.47}}                             & \multicolumn{1}{c}{0.40 $\pm$ 0.02}                                 & 0.37 $\pm$ 0.02                                      \\ \bottomrule
\end{tabular}
\caption{Performance comparison for regression models with 95\% confidence intervals (cross-validation scores for root mean squared error and test scores for weighted absolute percentage measure).}
\label{tab:Regression_results}
\end{table}

Given that the inclusion of alternative predictors did not substantially improve the models' performance, we followed a parsimonious principle and selected the best models using only financial features as predictors. This fact was an essential insight because it indicated that, in this specific case, customers' behaviors related to credit cards were sufficient to predict the future outstanding balance; thus, for certain tasks, fintech companies may encounter similar challenges as traditional banks, without necessarily enjoying an advantage due to their additional data collection practices. This result contrasts with that obtained by \citet{roa2021super}, who observed a moderate gain in the prediction of default using alternative data. It is worth noting that the design and construction of this simulator could also be undertaken by the traditional banking industry, which relies solely on conventional data features and may not have alternative data sources.

Finally, the evaluation measures over the ensemble model (i.e., applying the classification model and then the regression model) showed a test RMSE equal to 132.32 USD (95\% CI: [115.27, 149.37]). This corresponds to 49\% of the test target standard error, and a reduction of 50\% compared with the model in which the mean of the training target is the prediction value whose test RMSE is 268.32 USD. Finally, the overall WAPE is 0.54 (95\% CI: [0.49, 0.58]), which is superior to the WAPE of 1.36 obtained when using the mean of the training target.

\begin{figure}[htbp]
\centering
	\includegraphics[width=0.6\textwidth]{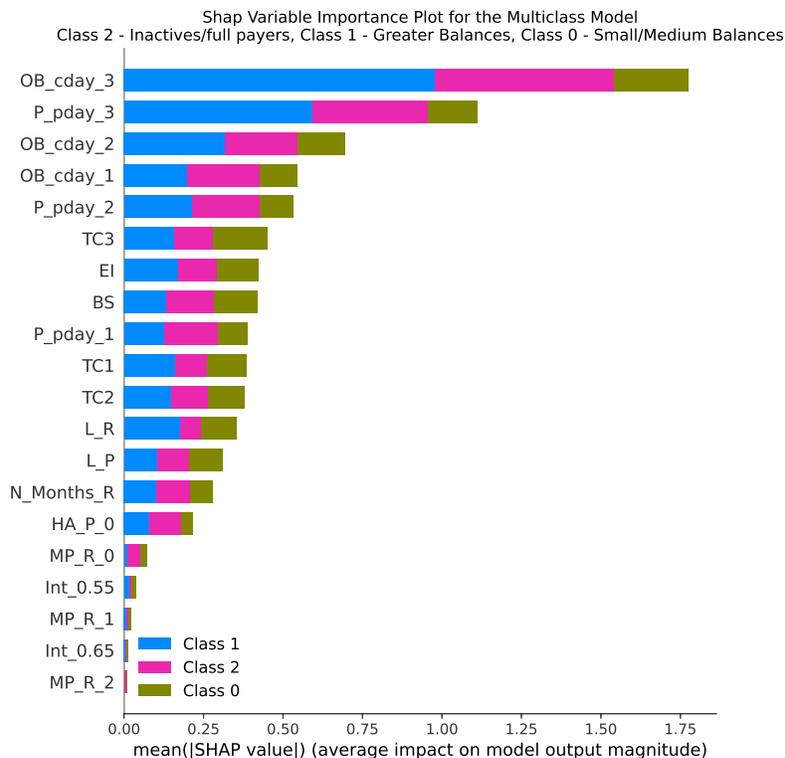}
	\caption{Global variable importance for the classifier.}
	\label{fig: GlobalImportance_Classifier}
\end{figure}

On the other hand, because the models we implemented were non-linear, we used SHapley Additive exPlanations (SHAP) to interpret how the predictors had affected their outputs \citep{lundberg2017unified}. Figure~\ref{fig: GlobalImportance_Classifier} presents the global feature importance obtained for the classifier model. From this result, we concluded that outstanding balance and payment in the last month were the most important predictors for determining classes 1 (large amount of debt) and 2 (inactive or full payers). In contrast, last outstanding balance and total spending consumption in the last month had the most significant predictive power for determining class 0 (small or medium balances). The credit limit in the retrospective window had moderate predictive power, and we observed that it could be used more effectively to determine class 1 and class 0 rather than class 2. The new credit limit imposed after taking action (increase or maintain) had lower importance overall than the older one; however, its importance is nearly equal for all classes.

\begin{figure}[htbp]
     \centering
     \begin{subfigure}[b]{0.45\textwidth}
         \includegraphics[width=\textwidth]{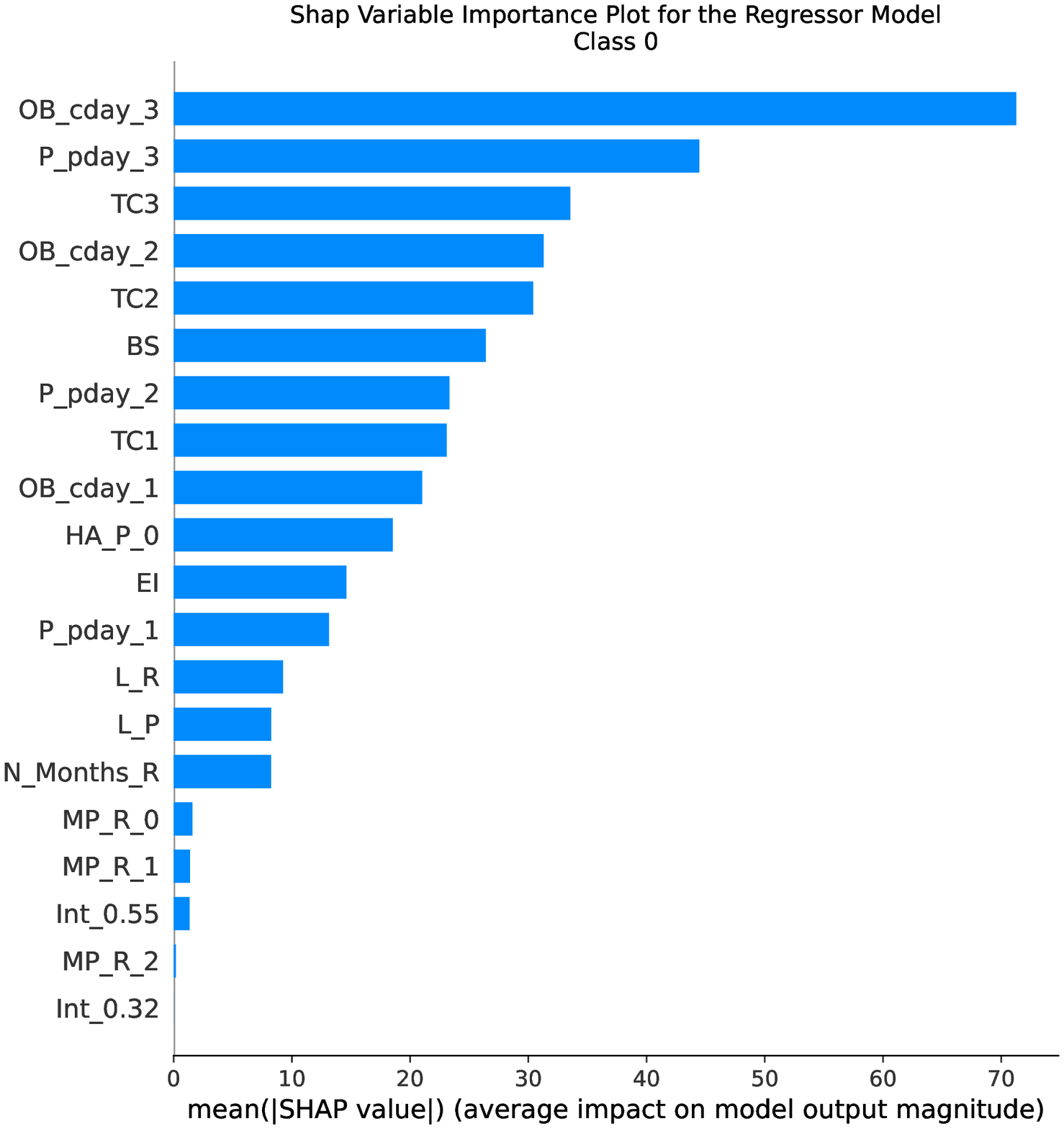}
         \caption{Regressor-class 0.}
         \label{subfig:Global variable importance class 0}
     \end{subfigure}
     \hfill
     \begin{subfigure}[b]{0.45\textwidth}
         \includegraphics[width=\textwidth]{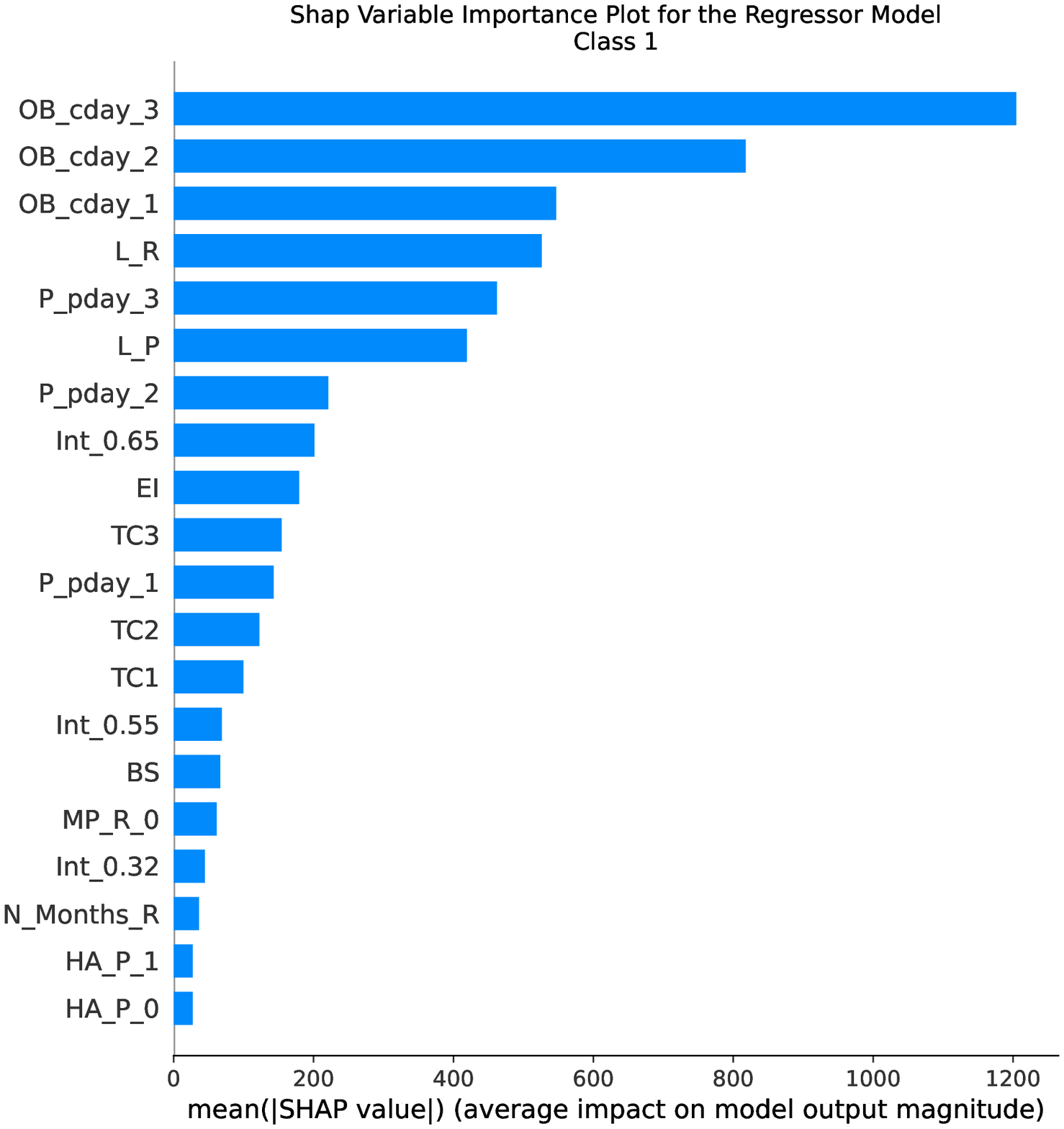}
         \caption{Regressor-class 1.}
         \label{subfig:Global variable importance class 1}
     \end{subfigure}
     \caption{Global variable importance for regressors.}
     \label{fig: Global variable importance regressors.}
\end{figure}

Graphs of global variable importance for the regressor models given the balance type are presented in Figures~\ref{subfig:Global variable importance class 0}
and~\ref{subfig:Global variable importance class 1}. For a prediction that the balance amount would be of class 0 (predicting a small or medium balance), the most important variables were the last outstanding balance and the last payment, followed by the total consumer spending in the last month of the observation period. The interest charged had little predictive relevance. For a prediction that the balance would be of class 1 (predicting a balance greater than 75.81 USD), the variables with the highest predictive power were all three previous outstanding balances and the credit limit before taking the action. 
\newpage
\subsection{RL training algorithms}
\label{subsec:RLtraining_result}

After creating the simulator, we use it to train the Double Q-learning algorithm; because we wanted our agent to maximize the long-term rewards we set $\gamma =1$. Additionally, we determined $CCF$ as the sample mean of the individuals' credit conversion factor given by~\eqref{eqn:CCFd} for the defaulters' accounts over all the credit card history up to December 2021 \citep{engelmann2011basel}. Subsequently, we discretized the states to use tabular RL methods. From the states defined in~\eqref{eqn:state}, we made uniform partitions for $\overline{UR}_n$,$\overline{PR}_n$, and $\overline{CR}_n$, each from 0\% to 100\% with 5\% intervals. For the current credit limit, given that customers with credit limits below 1,000 USD are more common, we used a finer grid than credit limits above this number. Finally, the $\Delta Provisions$ variable was divided by every 1\% of the possible total change in provisions if all customers had received the increment of $\beta$ in their current credit limit.

\begin{figure}[h]
     \begin{subfigure}[b]{0.45\textwidth}
         \includegraphics[height=5.4cm, width=\textwidth]{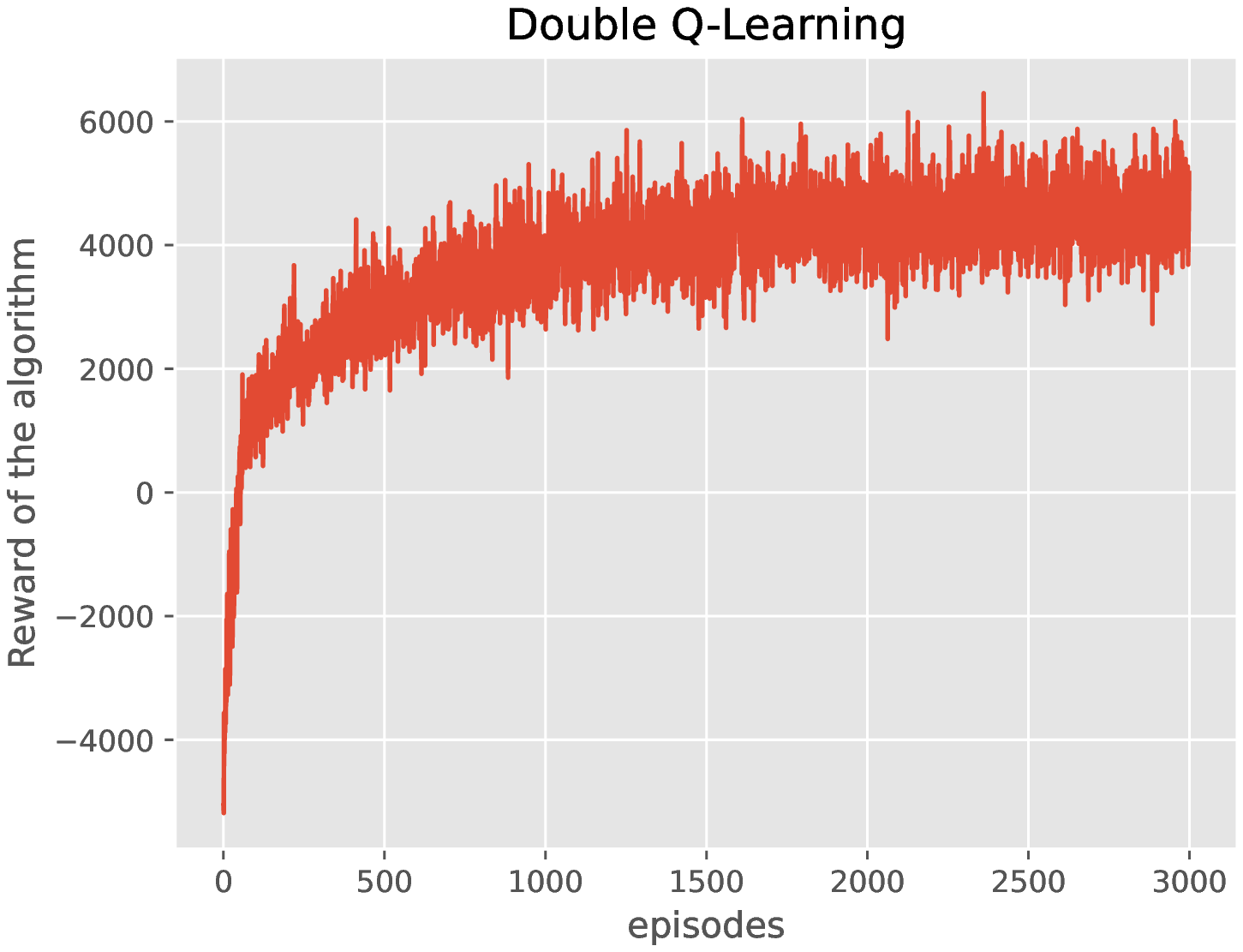}
         \caption{Raw Learning process.}\label{subfig:CompleteReward}
     \end{subfigure}
     \hfill
     \begin{subfigure}[b]{0.45\textwidth} 
         \includegraphics[height=5.4cm, width=\textwidth]{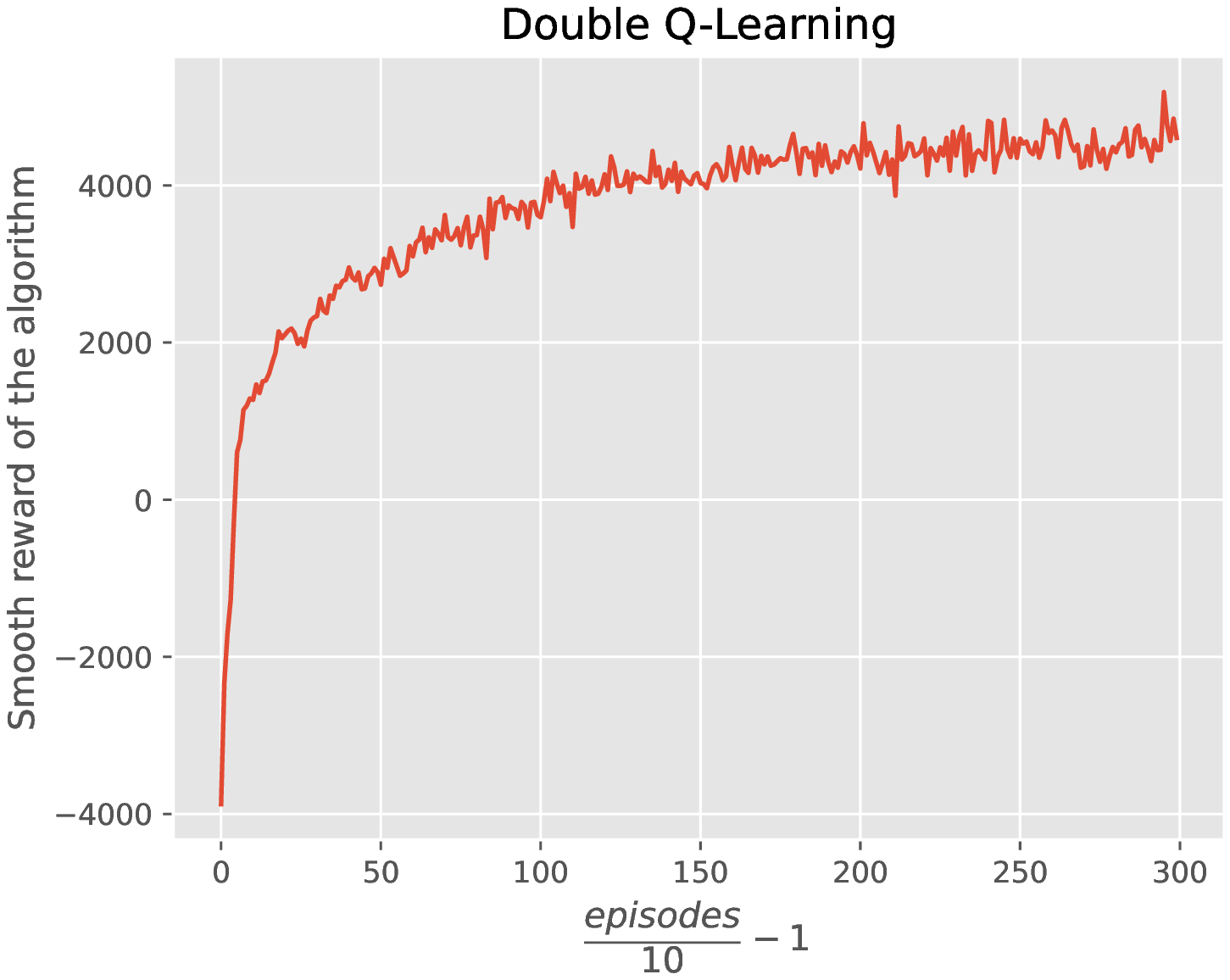}
         \caption{Smoothed learning process.}\label{subfig:SmoothenReward} 
     \end{subfigure}
         \hfill
         \begin{subfigure}[b]{0.47\textwidth}
         \ \ \ \ \includegraphics[height=5.4cm,width=\textwidth]{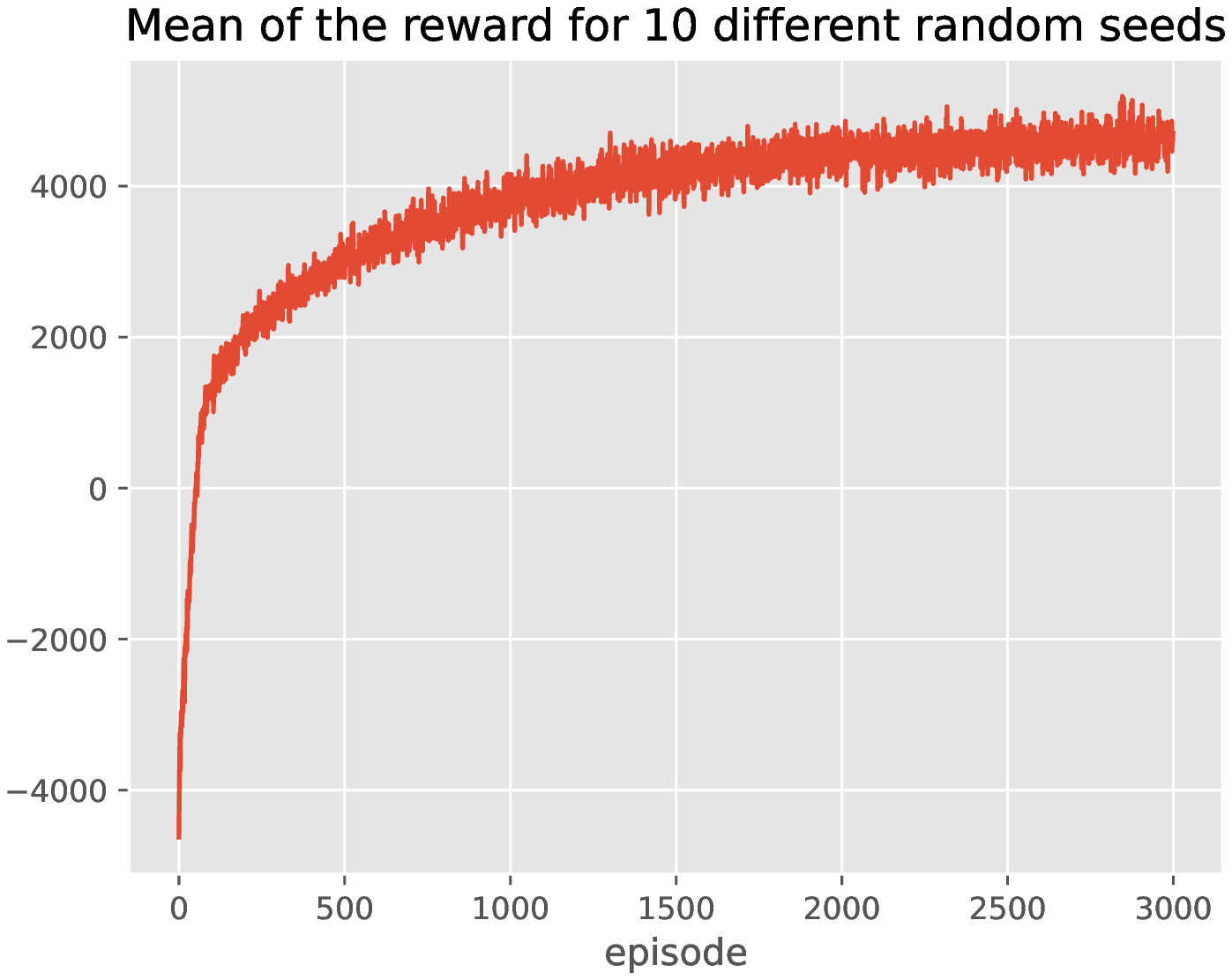}
         \caption{Mean Learning process over 10 random seeds.}\label{subfig:Mean_RandomSeeds}
     \end{subfigure}
     \hfill
     \begin{subfigure}[b]{0.47\textwidth}
         \ \ \ \ \ \ \ \ \ \includegraphics[height=5.4cm, width=\textwidth]{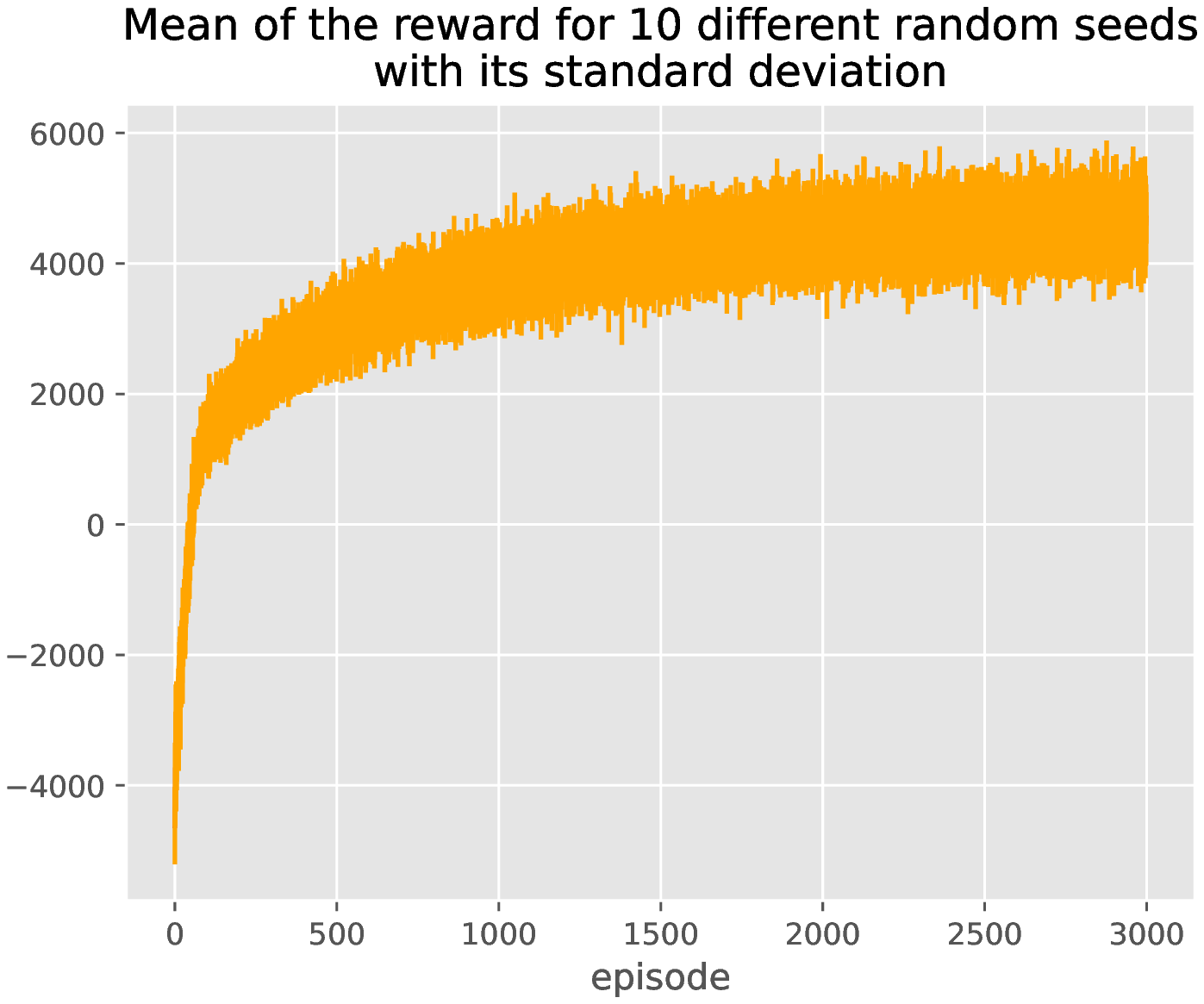}
         \caption{Mean and standard deviation of the learning process.}\label{subfig:Mean_Std_RandomSeeds}
     \end{subfigure}
     \caption{Learning process.}\label{fig:RL_learning_process}   
\end{figure}

The next step involved training the Double Q-learning algorithm. First, we searched for its hyperparameters using a grid-search procedure over the $\epsilon$ and $\alpha$ parameters, the results of this training strategy could be found in the~\ref{sec:appendix}. Subsequently, we trained the Double Q-learning algorithm with the best found hyperparameters, for this specific case, they were $\epsilon=0.1$ and $\alpha=10^{-2}$. It is relevant to highlight that in every episode, we permuted the rows of the training set to avoid any bias due to the order of the data. The raw results of this training process are shown in Figure~\ref{subfig:CompleteReward}, while the smoothed moving average over 10 episodes is shown in Figure~\ref{subfig:SmoothenReward}. In addition, we performed a robustness check of the algorithm by training it over 10 different runs and changing the random seeds. Figure~\ref{fig:RL_learning_process} also presents a plot of the means of the learning processes over 10 different runs (Figure~\ref{subfig:Mean_RandomSeeds}) and their means and standard deviations (Figure~\ref{subfig:Mean_Std_RandomSeeds}). 

Furthermore, because the company had followed an ad-hoc strategy until the moment at which the data was provided, we compared the Double Q-learning reward process against the rewards given by the use of baseline policies, including those of acting randomly (selecting the increases randomly), increasing limits for all customers, and maintaining the same credit limits.

\begin{figure}[h]
\centering
	\includegraphics[width=0.8\textwidth]{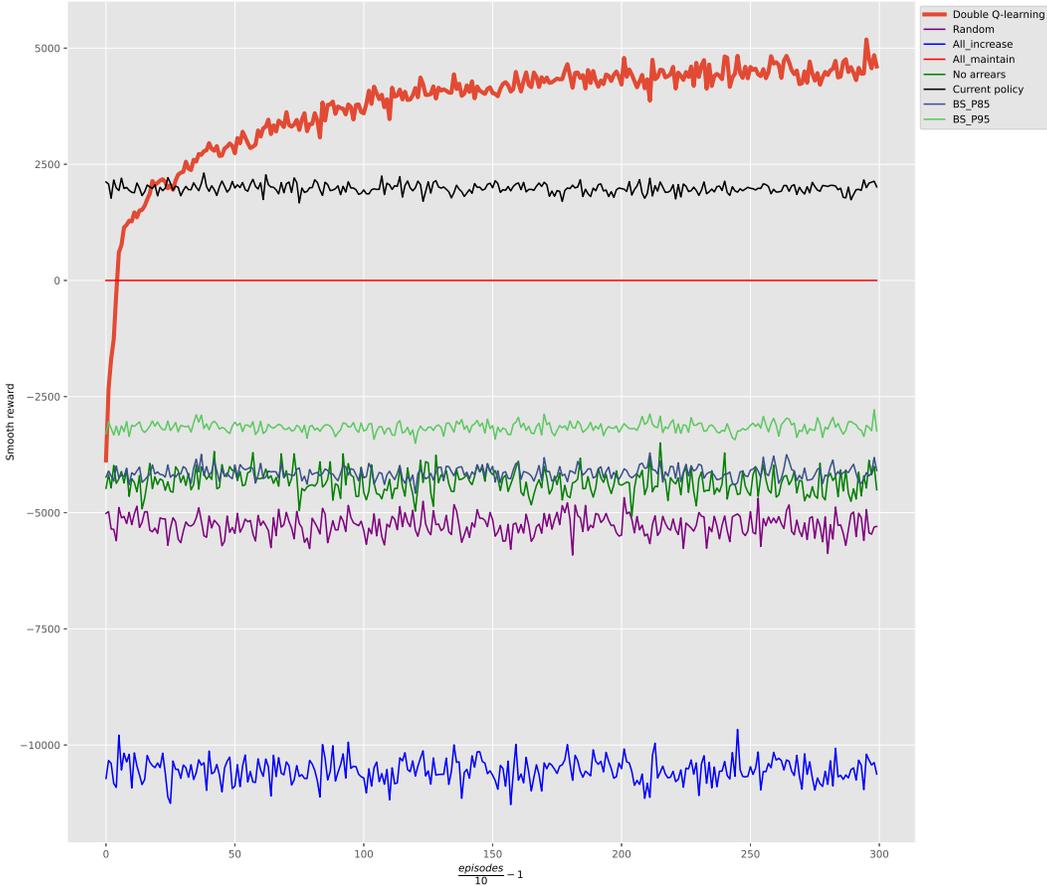}
	\caption{Comparison of different strategies for adjusting credit limits.}
	  \label{fig:Comparison_strategies}
\end{figure}

Figure~\ref{fig:Comparison_strategies} shows that the Double-Q-agent (i.e., the Double Q-learning algorithm) outperformed other policies in our synthetic trials. Since for the problem of credit limit adjustment there is no standard method to follow, as opposed to traditional questions such as the prediction of default probability in which logistic regression is extensively used in industry, we compared our results with the following strategies:

\begin{itemize}
    \item \textbf{Random:} This procedure consisted in selecting the increases in an indiscriminate way.
    \item \textbf{All Increase:} In this case we assumed increase action for all the card holders.
    \item \textbf{No arrears:} Consisting in increase to all the customers that did not have any arrears in the retrospective window.
    \item \textbf{Current policy:} To generate this policy, due to the past variations in the company's strategy, we trained a propensity score model to decide which customer should receive an increase based on the history of the decisions made in the given data.
    \item \textbf{BS\_P85:} This approach consisted in increasing the customers with Bureau scores greater than the 85 percentile of the available scores.
    \item \textbf{BS\_P95:} This strategy is similar to the previous one but considers the 95 percentile.
\end{itemize}

\begin{figure}[h]
\centering
	\includegraphics[width=0.9\textwidth]{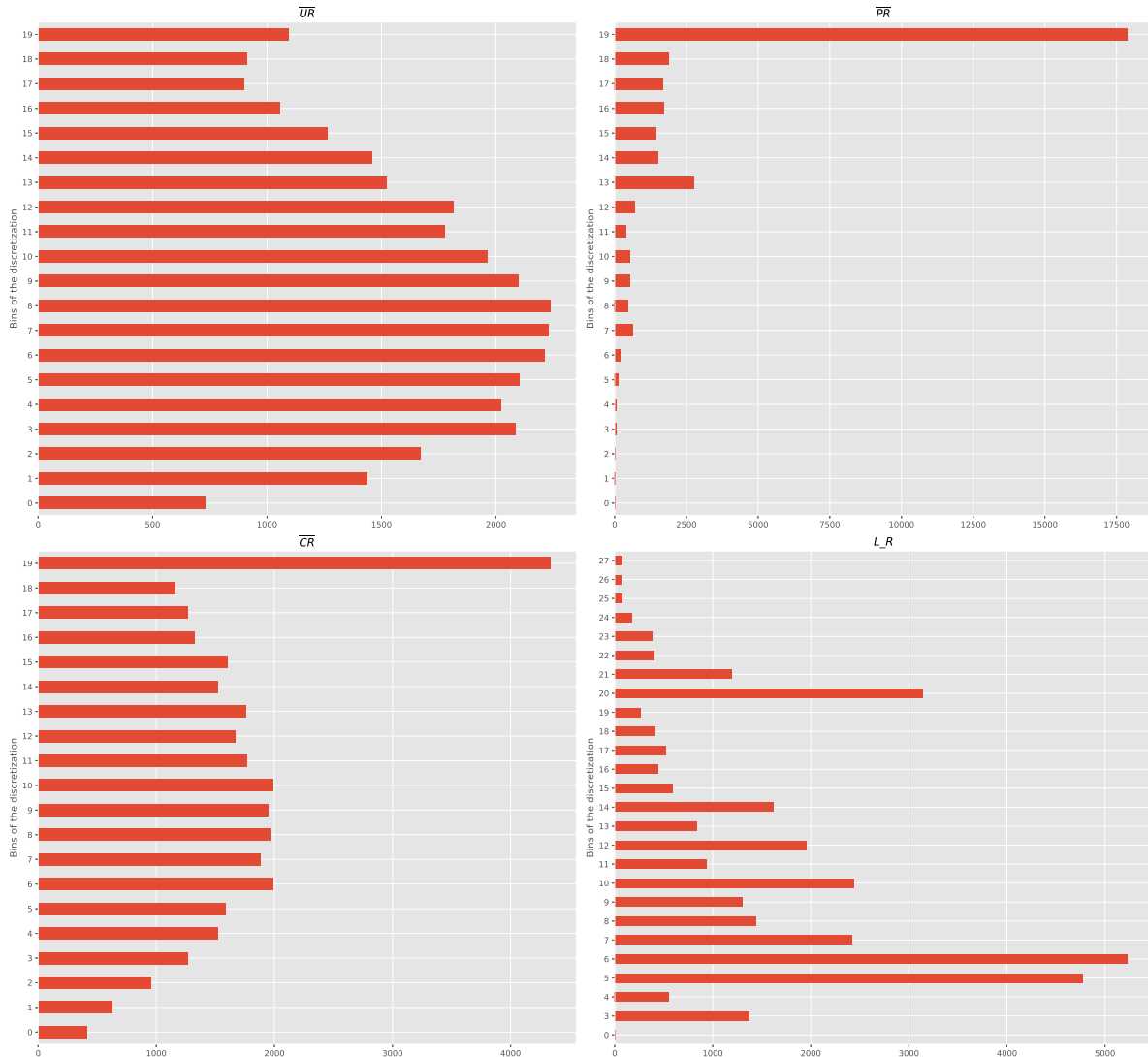}
	\caption{Histograms monthly average utilization, payment, spending consumption rate, and current limit.}
	  \label{fig:first4}
\end{figure}

\begin{figure}[h]
     \begin{subfigure}[b]{0.45\textwidth}
        \includegraphics[height=9cm,width=\textwidth]{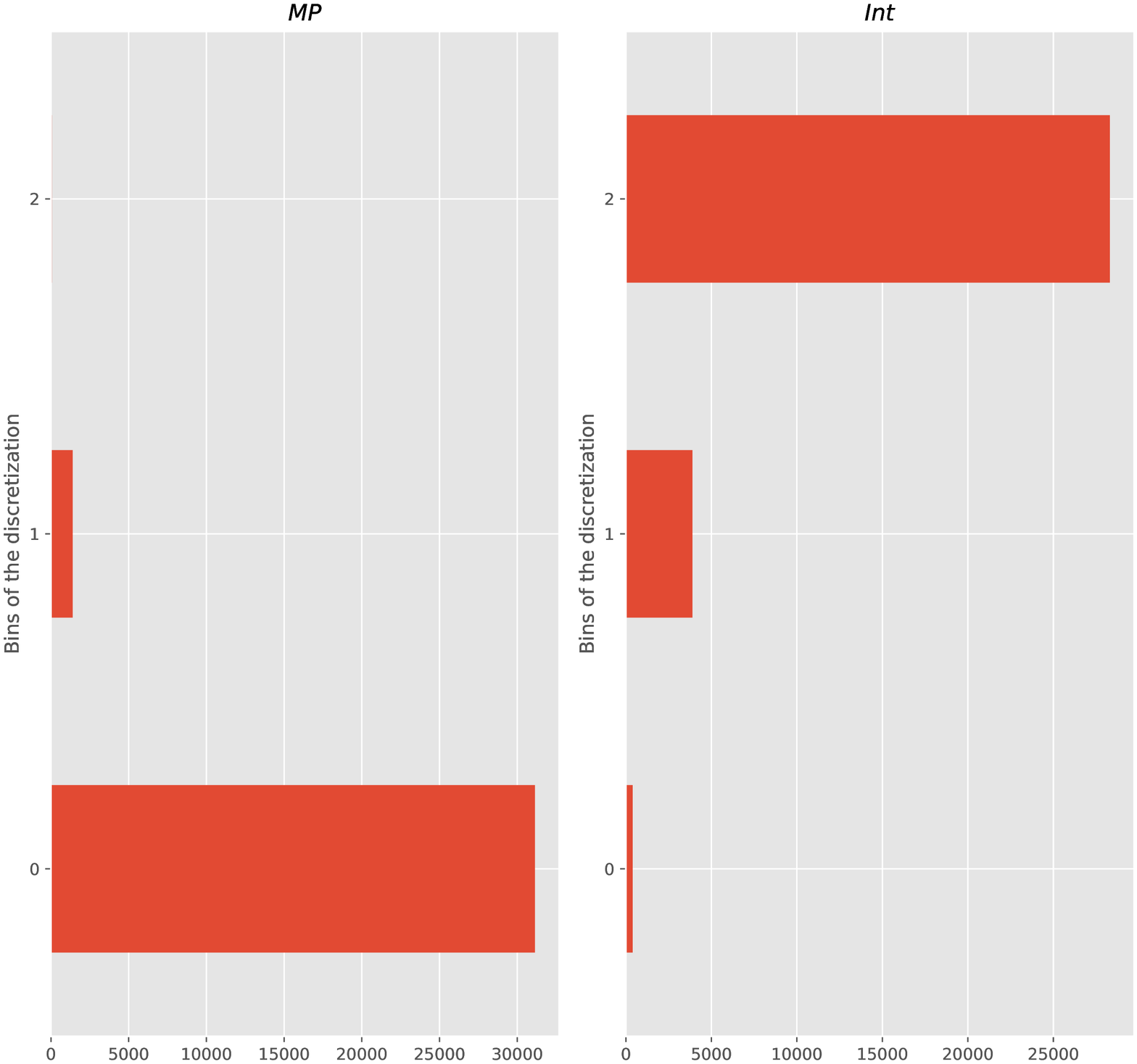}
	\caption{Histograms number of no payment and charged interest}
	  \label{fig:Second}
     \end{subfigure}
     \hfill
     \begin{subfigure}[b]{0.45\textwidth}
         \includegraphics[height=9cm,width=\textwidth]{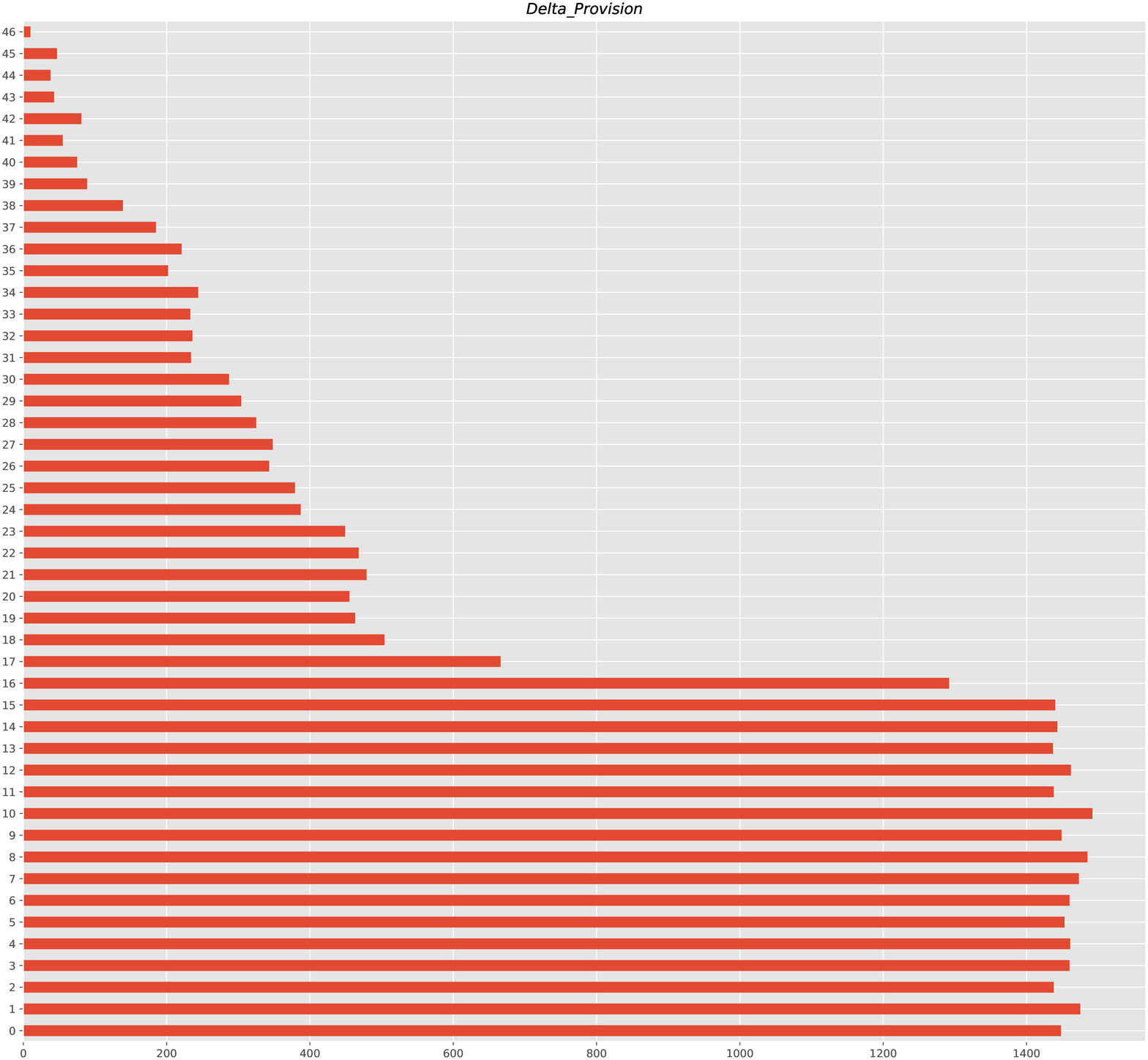}
	\caption{Histogram change in the provisions}
	  \label{fig:last}
     \end{subfigure}
     \caption{Histograms number of no payments, interest and provisions change.}
\end{figure}

To deduce the generated policy, we applied a Q-greedy policy to one of the resulting Q-tables, that is, to choose the action with the maximum $Q$-value for each customer, depending on their current state. For this procedure, we assessed the data frame in the order given by the last episode of the training process. Consequently, with this method, our agent recommended that 19\% of the customers should receive an increase in their credit limit. Figures~\ref{fig:first4},~\ref{fig:Second}, and~\ref{fig:last} present histograms of the discretized variables for the customers who were assigned an increase.

Figure~\ref{fig:first4}, shows that our agent preferred increasing the current credit limit to customers with a monthly utilization rate from 20\% to 40\% and that it very rarely increased it for customers with utilization rates below 5\%. In addition, the agent mainly suggested increases for customers with payment rates of at least 35\% and preferred customers with  payment rates above 90\%; this decision could be understood as expecting that with the increase of the limit, a full-paying customer would begin to consume more and thus begin paying more interest if they were not to pay their balance in full. Regarding spending consumption, customers with total consumer spending between 30\% and 50\% or greater than 90\% were more likely to receive an increase. We interpret this decision as assigning a higher credit limit to people spending substantially through the credit card but exhibiting positive payment behavior, which we had observed previously regarding payment rate. Finally, the higher the current limit, the less likely the customer is to be selected for a limit increase; this could be explained by the fact that most of these customers did not have an adequate rate of use or spending consumption (according to what the model considers optimal behavior) to justify an increase in their current limit.

As shown in Figures~\ref{fig:Second} and ~\ref{fig:last}, as expected, our agent rarely recommended increasing limits for customers with two delays in the last 3 months, and it predominantly offered increases to customers without any late payments; in addition, it preferred to offer increases to customers who were being charged a higher interest rate. Finally, the model tolerated a total increase in provisions equal to 46\% of the provisions if the company had increased the limit to every up-to-date customer in the portfolio. In particular, the provisions changed uniformly until arriving at 14\% of the total provisions possible; subsequently, the offers for increases became more restrictive.

It is important to emphasize that the results presented are based on the application of our simulator-driven methodology. As a result, these findings hold significant value as a trigger for testing innovative methodologies, such as RL, to address critical issues not only within the traditional banking industry but also among fintechs and super-apps offering financial products. By relying on objective methodologies rather than expert-driven approaches, we have the potential to enhance fairness in access to and conditions of financial products, particularly in the realm of credit cards. These products are an essential part of modern financial life, with over 175 million people in the United States reporting having one, representing more than half the population, owning credit cards according to the \citet{cfpb2021}, while in the LATAM region, approximately 20\% of the population utilizes them \citep{WEF2022}. Consequently, these methodologies can be adopted and tested by all these companies, regardless of the interface they use to engage with their customers, be it physical or virtual. Although fintechs and super-apps are data-superior, they are also competing with the long-term customer relationships that traditional banks have constructed over the years \citep{valverde2020financial}. Thus, employing more objective methods can contribute to healthier competition and an improved quality of product offerings for clients.

\section{Deployment}
\begin{figure}[h]
\centering
	\includegraphics[width=0.9\textwidth]{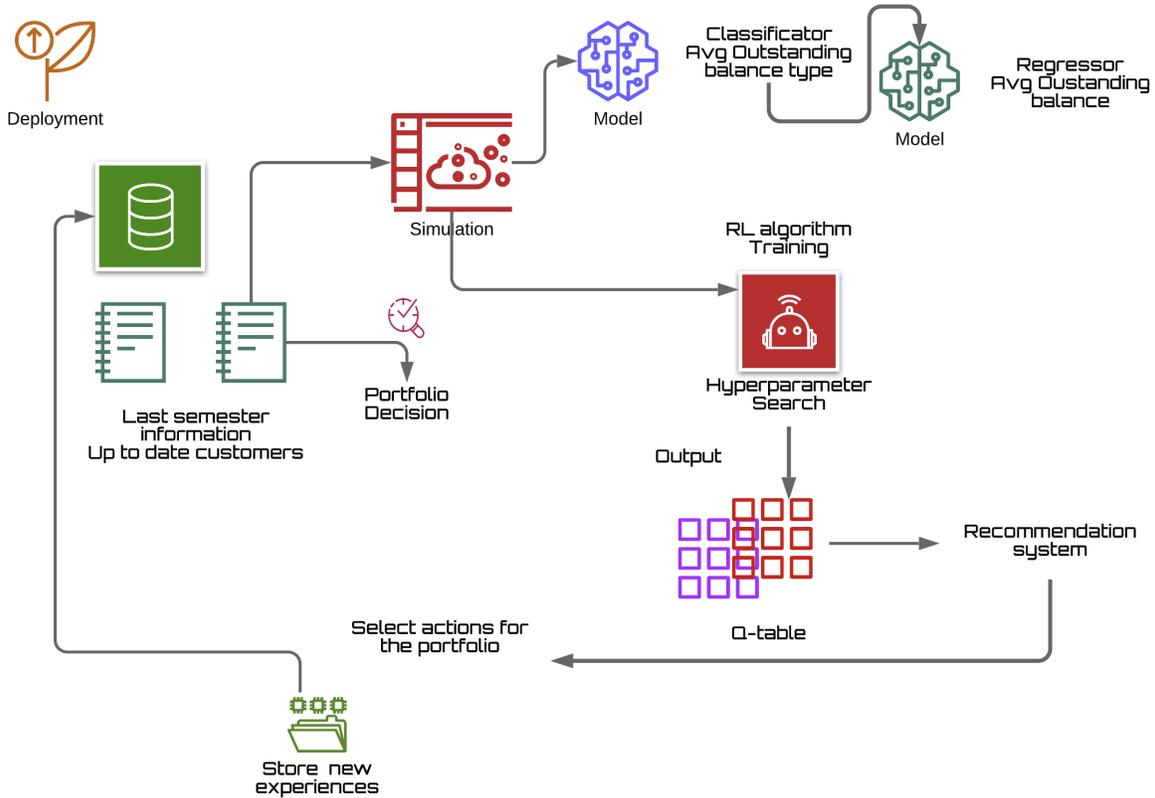}
	\caption{Methodology for deployment.}
	  \label{fig:Deployment}
\end{figure}

As the final goal of these efforts is to be able to apply this methodology in practice, Figure~\ref{fig:Deployment} presents the workflow of the process required to deploy the model suite.

First, the most recent 6 months of information regarding customers' behavior related to the use of their credit card should be used to create the simulator, in which the target variable should be the monthly average outstanding balance at the payment date, taking the last 3 months into consideration. For the studied period, the best model is a two-stage model consisting of one XGBoost multiclassifier and a mixture of regression models, including only financial information as predictors. Over time, this process will need to be changed to account for model updates because after 6 months of new data are available, customers may change their behavior, resulting in the need to retrain the model(s). This retraining phase also serves to evaluate whether the incorporation of newly available alternative data can enhance the performance of the model(s). Once the models have been trained, they will act as the base for creating the environment simulator. During offline RL training, the hyperparameter search, discretization of the space, and $CCF$ calculation will be required. 

Finally, after selecting the ideal hyperparameters and training the RL model, the output must be used to provide decision support regarding credit limit adjustment, specifically to increase the credit limit by a factor $\beta$ or to maintain the limit at its current level. After applying this policy, its outcome in terms of future balances, utilizations, and other metrics should be stored, which will help to produce a more robust and accurate simulator in the following 6 months. Here, it might be expected that consideration of longer time windows would produce less accurate prediction results; however, this possibility could be assessed when the credit card portfolio has developed further.

\section{Conclusions}

Our aim in this research was to characterize the adjustment credit limit decision as an RL problem, evaluate the predictive impact of the use of alternative data generated by the use of a super-app (which also offers financial products) over balance prediction, and finally test, through synthetic experiments, if the policy generated by the RL algorithm outperforms baseline policies such as increasing the credit limit to all the customers that are up to date, maintaining unchanged the portfolio’s credit limits, or acting randomly. In line with these goals, in this article, we have presented a methodology that uses a traditional RL algorithm to adjust credit limits. Namely, we formulated the problem within the RL framework by defining all the elements in this theory: agent, environment, states, and reward function. In contrast with previous works, we found that it is crucial that the reward function accounts for not only expected revenue but also expected losses, including the concept of the CCF to estimate future exposure at default. Indeed, in different experiments, we found that including the CCF factor was a major determinant in the final learning process of this RL agent. As opposed to the deterministic environments commonly used in RL, we used a customer-oriented, stochastic environment predicted using ML. This allowed us to formulate a strategy for the construction of the environment simulator that we used successfully to train the RL algorithm.

Secondly, our analysis revealed that incorporating alternative data sourced from the Latin America super-app does not consistently result in a significant improvement in the performance of our predictive models for forecasting both the type and amount of future balances. As a result, for this particular problem and the specific alternative data associated with app usage, the company may encounter similar challenges to those faced by traditional banks. This emphasizes the idea that, in certain aspects of data-driven financial risk management, fintech companies may not always diverge significantly from traditional banking practices. Finally, in our simulated experiments, our RL agent outperformed other strategies in the context of credit limit modifications. This finding not only underscores the potential of our proposal but also provides a compelling basis for further exploration of this innovative methodology in other real-world scenarios. We have also provided a general overview of the process necessary to deploy an RL-based limit-setting strategy to be executed every 3 months; it is essential to note that the agent can provide support for company decisions but that the final judgment should always be made in agreement with a human analyst involved in the process.

Although this methodology generates a recommendation for deciding to whom customers the company should increase the current credit limit, given the data that we had available, we mainly observed experiences with a fixed $\beta$ factor of increase. Therefore we could not test a more general framework for deciding between more than two actions. However, as with the time windows for measuring the impact of the actions, this could be investigated when the credit card portfolio is more mature. On the other hand, if there is a desire to explore the possibility of reducing credit limits, it's essential to recognize the high reputational risk associated with such actions. Even in the presence of adverse events, these decisions to lower credit limits may be perceived more as commercial choices rather than financial ones, a distinction that falls outside the scope of this work. 

As future work related to the development of a credit card portfolio, we aim to establish a more general framework in which the factor of credit limit increment is not unique and in which we can generate individualized recommendations. For this purpose, we will attempt to use counterfactual predictions given different amounts or percentages of increments (\textit{``treatments"}) using deep learning models. 

\section{Acknowledgement}
The authors gratefully acknowledge the support of the Natural Sciences and Engineering Research Council of Canada (NSERC) [Discovery Grants RGPIN-2020-07114 and RGPIN-2019-06586]. This work was funded, in part, with funding from the Canada Research Chair Program [CRC-2018-00082], and was enabled, in part, by support provided by Compute Ontario (\url{https://www.computeontario.ca}) and the Digital Research Alliance of Canada (\url{https://alliancecan.ca}).

\appendix
\section{RL Hyperparameter search}
\label{sec:appendix}
The $\epsilon$ parameter is responsible for the exploration-exploitation trade-off. The $\alpha$ is the learning rate, which measures how much weight to assign to new experiences. To determine the grid, we first conducted preliminary experiments with $\epsilon \in (0, 1)$, observing that the runs with $\epsilon$ greater than 0.2 had performed poorly in terms of average rewards. We then used the following grid, in which the $\alpha$ parameter had taken more possible values:
\begin{align*}
\epsilon &\in \left\{ 0.05, 0.1, 0.15\right\}\\
\alpha &\in \left\{ 10^{-6},10^{-5},10^{-4},10^{-3},10^{-2}, 10^{-1} \right\},
\end{align*}

\begin{figure}[htbp]
     \centering
     \begin{subfigure}[b]{0.3\textwidth}
         \includegraphics[width=\textwidth]{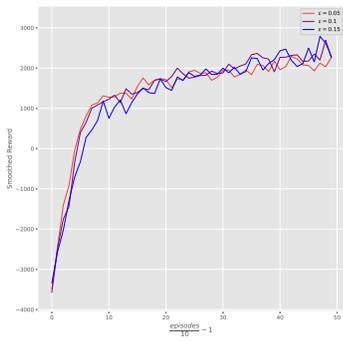}
         \caption{Training with $\alpha =10^{-1}$}\label{subfig:alpha0.1}
     \end{subfigure}
     \hfill
     \centering
     \begin{subfigure}[b]{0.3\textwidth}
         \includegraphics[width=\textwidth]{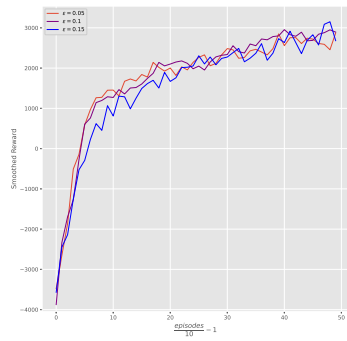}
         \caption{Training with $\alpha =10^{-2}$}\label{subfig:alpha0.01}
     \end{subfigure}
     \hfill
     \begin{subfigure}[b]{0.3\textwidth}
         \includegraphics[width=\textwidth]{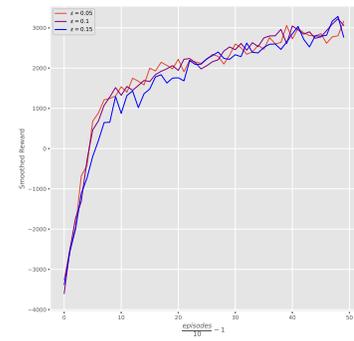}
         \caption{Training with $\alpha =10^{-3}$}
         \label{subfig:alpha0.001}
     \end{subfigure}
     \caption{RL hyperparameter search.}\label{fig:RLhyperparameter_search}
\end{figure}

Figure~\ref{fig:RLhyperparameter_search} presents plots of the average reward over 10 episodes after training 500 episodes for various combinations of $\epsilon$ and $\alpha$ for the Double Q-learning algorithm. According to these results, using $\alpha = 10^{-1}$ was less effective than using $\alpha = 10^{-2}$, while for $\alpha<10^{-2}$ the results were very similar to those with $\alpha=10^{-2}$. Therefore, we selected $\alpha=10^{-2}$. The plots in Figure~\ref{subfig:alpha0.01}, show that $\epsilon=0.1$ produced the best result. 

\end{document}